\documentclass[usenatbib, useAMS]{mn2e}

\usepackage{natbib, graphics, graphicx, epsfig, color, times, amssymb, amsmath, verbatim,amsbsy}

\usepackage{times} 
\usepackage{amsmath}

\renewcommand{\vec}[1]{ \mathbf{#1} }

\title[Constructing equilibrium N-body galaxy models]{An iterative
  method for the construction of N-body galaxy models in collisionless
  equilibrium}

\author{Denis Yurin, Volker Springel}

\author[D. Yurin \& V. Springel]{\parbox{18.5cm}{
Denis Yurin$^{1,2}$ and
Volker Springel$^{1,2}$}\vspace{0.2cm}\\
$^1$Heidelberg Institute for Theoretical Studies, Schloss-Wolfsbrunnenweg 35, 69118 Heidelberg, Germany\\
$^2$Zentrum f\"ur Astronomie der Universit\"at Heidelberg, ARI, M\"onchhofstr. 12-14, 69120 Heidelberg, Germany\\
}

\begin{document}

\maketitle
\begin{abstract}
  We describe a new iterative approach for the realization of
  equilibrium N-body systems for given density distributions. Our
  method uses elements of Schwarzschild's technique and of the
  made-to-measure method, but is based on a different
  principle. Starting with some initial assignment of particle
  velocities, the difference of the time-averaged density response
  produced by the particle orbits with respect to the initial density
  configuration is characterized through a merit function, and a
  stationary solution of the collisionless Boltzmann equation is found
  by minimizing this merit function directly by iteratively adjusting
  the initial velocities. Because the distribution function is in
  general not unique for a given density structure, we augment the
  merit function with additional constraints that single out a desired
  target solution.  The velocity adjustment is carried out with a
  stochastic process in which new velocities are randomly drawn from
  an approximate solution of the distribution function, but are kept
  only when they improve the fit. Our method converges rapidly and is
  flexible enough to allow the construction of solutions with third
  integrals of motion, including disk galaxies in which radial and
  vertical dispersions are different. A parallel code for the
  calculation of compound galaxy models with this new method is made
  publicly available.
\end{abstract}

\begin{keywords}
stars: kinematics and dynamics -- methods: numerical -- galaxies: haloes -- galaxies:
kinematics and dynamics -- galaxies: structure
\end{keywords}

\pagerange{\pageref{firstpage} -- \pageref{lastpage}}
\pubyear{2014}

\label{firstpage}

\section{Introduction}  \label{sec:intro}

The large number of stars and dark matter particles in galaxies and
galaxy clusters makes them essentially perfect collisionless systems.
Their dynamics is hence described by the collisionless Boltzmann
equation, coupled to self-gravity through Poisson's equation. Relaxed
systems correspond to stationary solutions of these equations, and
much of the field of galactic dynamics is concerned with understanding
different aspects of these solutions \citep[see][for an excellent
exposition]{BinneyTremaine2008}. This is particularly important for
using observational probes of kinematics to infer, for example,
something about the underlying density distribution.

Numerical N-body simulations have become a primary work-horse to study
collisionless systems, both in stationary and dynamic
situations. Prominent examples include the study of bar instabilities
\citep[e.g.][]{Athanassoula2002}, the formation of spiral waves
\citep[e.g.][]{Donghia2013}, or major and minor mergers of galaxies
\citep[e.g.][]{Barnes1992, Hernquist1995}.  They are also actively
used to study the response of disks to the bombardment by dark matter
clumps \citep[e.g.][]{KazantzidisI2008, Donghia2010}, or the radial
migration of stars caused by resonance scattering
\citep[e.g.][]{Sellwood2002}, and many more.

In carrying out numerical experiments targeting these questions, a
recurrent challenge is to construct suitable initial conditions.  One
usually requires them to be in a reasonably stable, approximate
equilibrium in the beginning, otherwise any subsequent dynamics may be
dominated or heavily contaminated by the specific out-of-equilibrium
state one started out with. Often, one has a relatively clear notion
of the density structure one wants to realize, but initializing the
particle velocities appropriately is a quite non-trivial problem. This
is because doing this {\em perfectly} requires knowledge of the full
distribution function (DF) of the system, or in other words,
availability of a stationary solution of the collisionless Boltzmann
equation.  However, such solutions are analytically known only for a
very limited number of density distributions.

There is hence significant demand to construct equilibrium solutions
numerically, not only for realizing N-body initial conditions but also
in the context of modelling observational data sets.  In the latter
case, finding such models is a main component of the
reverse-engineering process aimed at constructing self-consistent
three-dimensional systems that reproduce the observations. They can
then be examined in great detail, allowing insights into properties
that are not directly observable \citep{Cretton1999, Bosch2012}.

To our knowledge, there are presently mainly five different methods in
use for constructing such equilibrium models:

\begin{enumerate}
\item DF-based: For certain mass distributions, the distribution
  function (DF) can be analytically calculated or accurately
  approximated.  Unfortunately, this ideal case is not generally
  available for arbitrary density distributions.  The main problem is
  that we do not know the analytical form of the third integral of
  motion. In some cases it may be reasonably approximated, but this
  leads at best to nearly self-consistent solutions
  \citep{Kuijken1995, Widrow2005}.  Nevertheless, there are some
  useful classes of solutions known, for example for spherical
  galaxies \citep{Osipkov1979, Jaffe1983, Merritt1985,
    Hernquist1990}. However, because many real systems are not
  particularly close to any of these parameterized classes of systems,
  the approach is rather restrictive in practice.

\item Moment-based: Moments of the velocity distribution can be
  calculated or estimated with the hierarchy of Jeans equations. If
  one neglects higher order moments and assumes a functional form for
  the velocity distribution \citep[often taken to be Gaussian,
    e.g.][]{Hernquist1993, Springel1999} that reproduces the estimated
  moments, one obtains an approximate distribution function.  This
  method is quite general and can be applied to all mass
  distributions. Since the true velocity distribution function is
  usually close to a triaxial Gaussian for much of the mass of a
  system, the method typically produces systems that are roughly in
  equilibrium. But the crux is that this equilibrium is by no means
  perfect, and that it is hard to overcome this limitation within this
  method. Especially difficult are the central regions of galaxies;
  when the constructed ICs are evolved in time, one here typically
  finds density ripples propagating through the system while it
  relaxes to a true equilibrium state. This can interfere with the
  interpretation of numerical experiments, especially when they
  require particularly quiet ICs.

\item Orbit-based method: \cite{Schwarzschild1979} introduced a
  radically different approach to solve the problem. He suggested to
  integrate a wide variety of orbits in a given potential, and then to
  distribute the mass of the system over this orbit library such that
  the time-averaged density of the system becomes as close as possible
  to the one corresponding to the potential.  Finding the weights of
  each of these orbits defines a linear optimization problem with
  positive coefficients, which can be solved iteratively. The
  resulting weights then effectively define the velocity distribution
  function.  A practical problem with this method is that the size of
  the orbit library is severely constrained by the available memory.
  Moreover, the method is ill-conditioned in its basic form, something
  that needs to be cured by adding ad-hoc assumptions such as
  smoothness constraints or maximum entropy measures for the
  weights. Also, the velocity distribution functions constructed with
  this method are typically very noisy and may feature large
  jumps. One needs to smooth them, but the required level of smoothing
  is hard to define. Many attempts have been made to overcome these
  difficulties \citep[e.g.][]{Vandervoort1984, Jalali2011}.

\item Made-to-measure: Attempts to improve on Schwarzschild's method
  have resulted in a new technique where the orbit integration process
  and the mass/weight redistribution are combined. This
  `made-to-measure' technique makes the storage of a full orbit
  library unnecessary and therefore removes the memory barrier. But it
  still requires a smoothing procedure for the velocity
  distributions \citep{Syer1996, Dehnen2009}.

\item Guided-relaxation: Another class of methods exploits the fact
  that any isolated system left to itself tends to an equilibrium
  state. Knowing the target mass distribution we may try to directly
  relax to it by steering a system appropriately in the process. This
  guiding can be done by introducing an additional force,
  e.g.~adiabatic drag on the vertical components of the particle
  velocities in order to squeeze the system
  \citep{Holley-Bockelmann2001}, or we may restrict particle mobility
  such that the target density distribution is maintained and the
  systems evolves towards a self-consistent equilibrium state
  \citep{Rodionov2009}. A disadvantage of this approach is that it
  involves one of the other methods to create an initial state for the
  further relaxation. Also, there is only limited control on the
  outcome, making it, e.g., difficult to construct systems with a
  prescribed velocity anisotropy.

\end{enumerate}

The purpose of this article is to introduce a new, flexible approach
for the construction of compound N-body models of axisymmetric
galaxies in an essentially perfect equilibrium state. The method only
requires the specification of the density profiles of the different
components and a selection of the desired bulk properties of the
velocity structure, such as the degree of rotational support or the
ratio between radial and vertical velocity dispersion in the disk
plane. Our code then constructs an N-body system that is in
equilibrium and fulfills the imposed constraints on the velocity
structure. Implicitly, it hence also provides a solution for the full
3D distribution function. This is achieved for essentially arbitrary
axisymmetric density structure and by taking the mutual influence of
different mass components (if present) fully into account.  We argue
that the resulting flexibility and accuracy makes our approach an
attractive alternative compared with other IC generation methods in
the literature.

This paper is structured as follows. In Section~\ref{sec:method}, we
describe the basic methodology adopted in our method, which consists
of an iterative procedure to adjust the velocities of an N-body
realization of a galaxy model until the prescribed density structure
is maintained self-consistently under time evolution, and the imposed
velocity constraints are fulfilled. In
Section~\ref{sec:velconstraints}, we highlight how we specify velocity
constraints for different structural choices. They take the form of
second velocity moments which we determine through solutions of the
Jeans equations. We then specify in Section~\ref{sec:implementation}
various implementation details of our numerical methods as realized in
the {\small GALIC} code introduced here. Section~\ref{sec:galaxies} is
concerned with a brief description of the specific density profile
models currently implemented in this code; these are employed for a
suite of tests presented in Section~\ref{sec:testing}. Finally, we
conclude with a summary of our findings in
Section~\ref{sec:conclusion}.

\section{Methodology}  \label{sec:method}

If density profiles for all collisionless mass components of a galaxy
model are given, a random N-body realization of particle positions can
be easily created by interpreting the density distribution as a
probability field for a Poisson point process. But assigning suitable
velocities to the particles is the difficult step. Our idea is to do
this iteratively: Starting from some guess for the particle
velocities, we try to correct them such that the system becomes closer
to the desired equilibrium state. This is in some sense similar to
Schwarzschild's method, where one models the distribution function
from which the velocities are drawn through a set of weights, which
are then iteratively adjusted until a global merit function is
extremized. Differently from this technique, we however optimize the
velocities of each particle directly. This eliminates the explicit
orbit library of the Schwarzschild method, and all the restrictions
that come with it. Instead, the particles of our N-body model
themselves define the orbit set that is optimized. Importantly, this
set is free of any discreteness restrictions or potential biases due
to incompleteness of the Schwarzschild orbit library.

As basic merit function that is optimized we consider the difference
between the target density field and the actual density response
created by our N-body realization with the currently assigned initial
velocities. The density response is here defined as the time-averaged
density field of the N-body orbits, calculated in the static potential
of the target density distribution. For a steady state system, this
density response is supposed to be time invariant, and equal to the
initial density field. We can readily imagine several different
optimization schemes that adjust individual particle velocities
iteratively such that this difference becomes as small as possible,
for example multi-dimensional steepest decent.

Before discussing the details of our specific solution for this, it is
however prudent to consider two apparent conceptual problems with the
basic approach as outlined thus far. One is that the density structure
does not uniquely specify the velocity structure of an equilibrium
model, or in other words, there can be more than one steady-state
solution of the collisionless Boltzmann equation for a given density
structure. For example, for a spherically symmetric mass distribution,
one can have solutions where the velocity distribution is isotropic
everywhere, and the distribution function depends only on energy (the
`ergodic' case). But there are also solutions with an anisotropy
between radial and tangential motions.  Furthermore, one can also have
many different axisymmetric solutions that feature different degrees
of net rotation.

It is hence not clear to which equilibrium solution our adjustment
scheme would converge when only the density response is
optimized. This ambiguity can be lifted by making a selection for the
desired type of solution one wants to obtain, and to suitably
incorporate this constraint in the merit function. For example, one
may request to obtain an anisotropic solution with a certain
prescribed ratio of radial and tangential velocity dispersions.  We
can then augment our density based merit function with further
conditions that enforce this velocity structure.

A second problem, of perhaps somewhat lesser importance, is the
possibility of overfitting individual particle velocities. In the
continuum limit of a collisionless system, individual particles are
completed uncorrelated from each other. An iterative optimization
approach will however always adjust a particle's velocity given the
current realization for positions and velocities of all other
particles. This can in principle introduce undesired correlations
between particles. Related to this, one may easily end up in an
unfavorable local minimum of the merit function.  We largely eliminate
this effect by using a special optimization strategy in which new
values for the velocity of a given particle are not searched in the
vicinity of the current velocity, but rather globally in a random
fashion, completely independent of the particle's current velocity.

In the following, we first discuss our formalism for determining the
density response of a particular realization and for measuring its
goodness of fit through a merit function. We then extend the
discussion to merit functions for the velocity moments, and present
our approach for optimizing both of them concurrently.

\subsection{Density merit function}

Consider a collisionless N-body system with $N$ particles, initial
coordinates $\vec{\hat{x}}_i$, and initial velocities
$\vec{\hat{v}}_i$. We assume that an initial density distribution
$\rho_0(\vec{x})$ is given, which can be used to create a
realization of the coordinates $\vec{\hat x}_i$ by random
sampling. Determining the $\vec{\hat{v}}_i$ is more complicated,
however; we want to initialize them such that a stationary solution of
the collisionless Boltzmann equation is obtained where the particles
move in a given, stationary gravitational potential
$\Phi(\vec{x})$. In other words, the collection of particles should
move such that the density field they create is time invariant and
identical to the initial density distribution. If this is achieved,
the gravitational field can then also be chosen self-consistently as
the one created by the mass distribution itself (plus additional
contributions by other mass distributions, if desired), such that one
obtains a stationary self-gravitating solution of the Poisson-Vlasov
system.

The density field created by the particles of our system can be
formally expressed through a superposition of Dirac delta functions:
\begin{equation}
\rho(\vec x, t, \vec{\hat v}_1, \ldots, \vec{\hat v}_N) = \sum^N_{i=1}
m_p~\delta(\vec x_{\rm orbit}(\vec{\hat x}_i, \vec{\hat v}_i, t) -
\vec{x}),
\end{equation}
where the function $\vec{x}_{\rm orbit}(\vec{\hat x}', \vec{\hat v}',
t)$ describes the time-dependent orbit of a particle starting in the
phase-space point $(\vec{\hat x}', \vec{\hat v}')$. Note that in the
expression for the density field we have explicitly retained the
dependence on the initial values of the particles velocities (which we
have yet to determine), whereas the initial positions can be viewed as
fixed parameters.

Next, we define the time-averaged density response for the chosen
initial velocities as
\begin{equation}
\bar{\rho}(\vec{x}, \vec{\hat v}_1, \ldots, \vec{\hat v}_N) =
\lim_{T\to \infty}\frac{1}{T}\int_0^{T} \rho(\vec x,t, \vec{\hat v}_1,
\ldots, \vec{\hat v}_N) \, {\rm d}t .
\end{equation}
The best steady-state for the system can be defined as the smallest
possible difference between the time-averaged density and the initial
density field. To this end, we introduce an objective function
\begin{equation}
S(\vec{\hat v}_1, \ldots, \vec{\hat v}_N) = \int \left| \,
\bar{\rho}(\vec{x}, \vec{\hat v}_1, \ldots, \vec{\hat v}_N) -
\rho_0(\vec{x}) \, \right| \, {\rm d}\vec x ,
\label{Sdiff}
\end{equation}
which measures the $L_1$-norm of the difference between the two
fields. The linear weighting of the mass difference at a given
location is motivated by the source term in Poisson's equation, which
is ultimately what we want to keep constant as much as possible in a
steady state to avoid potential fluctuations that can modify particle
energies.

Therefore, the task to construct a best possible steady-state that has
a given density distribution is primarily about finding the $\vec{\hat
  v}_i$ such that the difference $S$ defined by equation (\ref{Sdiff})
reaches a minimum. Note that this can in principle be viewed as a
high-dimensional minimization problem with respect to the initial
velocities. This could, for example, be tackled with the method of
steepest decent. A direct adjustment of the velocities to minimize the
function $S$ is indeed the central idea we pursue in this paper,
yielding a novel scheme for constructing equilibrium solutions.  There
are however a number of obstacles that make such a minimization
non-trivial.

First of all, the problem needs to be somehow discretized, otherwise
we cannot meaningfully define a density field for a finite number of
particles. We will deal with this aspect in the remainder of this
subsection. A further conceptual problem, namely the non-uniqueness of
the obtained solutions, needs to be addressed as well.

Let's assume we discretize the space covered by our system in terms of
cells of volume $V_j$, indexed by $j$. The cells cover the volume but
they do not need to be of the same size (e.g. we may choose to use
adaptive logarithmic grids, as we do in practice). The merit function
(\ref{Sdiff}) can now be written as
\begin{equation}
S(\vec{\hat v}_1, \ldots, \vec{\hat v}_N) = \sum_j \left| \,
\overline{M}_j(\vec{\hat v}_1, \ldots, \vec{\hat v}_N) - M_j^0
\right|,
\end{equation}
where $\overline{M}_j$ and $M_j^0$ give the masses of the
time-averaged and the initial density field that fall into cell $j$,
respectively.  To determine $\overline{M}_j$, we add the time-averaged
contributions of the orbits of all particles to this spatial cell.
Since the trajectories of different points require different times to
saturate their impact on the common averaged density, it is
computationally more efficient to follow their orbits over
individually chosen time scales $T_i$.  We can then write
\begin{equation}
\overline{M}_j = \sum^N_{i=1} \int_{{\rm cell}\; j} {\rm d}\vec{x}
\int \frac{{\rm d}t}{T_i} \; m_p~\delta(\vec x_{\rm orbit}(\vec{\hat
  x}_i, \vec{\hat v}_i, t) - \vec{x}).
\end{equation}
This reduces to
\begin{equation}
\overline{M}_j = \sum^N_{i=1} {m_p}~ \tau^{\rm orbit}_j(\vec{\hat
  x}_i, \vec{\hat v}_i),
\end{equation}
where $\tau^{\rm orbit}_j(\vec{\hat x}_i, \vec{\hat v}_i)$ gives the
fraction of time an orbit starting in the given phase-space point
spends in cell $j$. The expected mass in the cell, $M_j^0$, is simply
given by
\begin{equation}
M_j^0 = \int_{{\rm cell}\; j} \rho_0(\vec{x}) {\rm d}\vec{x}. 
\end{equation}
We note that the above equations correspond to so-called nearest grid
point assignment of the current position of a particle to the
mesh. One can replace this with a higher-order assignment scheme if
desired, with the simplest possibility being clouds-in-cell
assignment\footnote{We actually use the latter in our implementation,
  even though the improvement relative to nearest grid point
  assignment is here minor.}.

A more important question concerns the choice of the spatial binning
scheme. There should be enough bins to resolve all relevant detail of
the density distribution, but the Poisson noise affecting
$\overline{M}_j$ due to the finite number of particles we use clearly
limits the minimum size of a bin that is reasonable. In order to make
the noise in each bin comparable, it is advantageous to choose the
bins sizes such that they contain roughly equal mass. We follow this
strategy by adopting a hierarchical adaptive binning scheme combined
with a logarithmic grid. We will describe this approach in full detail
in Section~\ref{secbinning}.

\subsection{Velocity dispersion merit functions}

As we discussed earlier, the requirement of a stationary density field
does not in general imply a unique solution for the distribution
function. For example, in an axisymmetric system, it would always be
possible to flip the signs of the azimuthal velocities to generate,
e.g., a system where all particles orbit around the $z$-axis with
positive $L_z$, or with negative $L_z$, or with any desired mixture of
the two. It is hence unclear in which minimum one ends up when one
would try to directly minimize $S$ with respect to the $\vec{\hat
  v}_i$.

In order to lift this ambiguity and make the solution more well
defined, we need to add additional constraints that drastically reduce
the acceptable solution space. We do this by invoking symmetry
assumptions about the velocity structure of the system. This then
allows solving the Jeans equations for the second velocity moments,
which we impose as a further optimization constraint.

In general, there are three first moments and three second (reduced)
moments of the velocity distribution function at every point. We will
here focus on axisymmetric systems and employ cylindrical coordinates
$(R, \phi, z)$. In a stationary system, we always have
$\left<v_R\right> = 0$ and $\left<v_z\right> = 0$. We shall now assume
that as part of specifying our desired target system, we provide
enough assumptions such that the three dispersions and the azimuthal
streaming can be calculated everywhere, i.e.  $\sigma_R^2 =
\left<v_R^2\right>$, $\sigma_\phi^2 = \left<(v_\phi -
\overline{v_\phi})^2\right>$, $\sigma_z^2 = \left<v_z^2\right>$, and
the mean azimuthal streaming $\overline{v_\phi} = \left< v_\phi
\right>$ can be considered to be known as a function of $(R,z)$.  How
we compute these quantities in practice for different cases will be
discussed in Section~\ref{sec:velconstraints}.

In order to impose these velocity moments as additional constraints on
the initial velocities $\vec{\hat v}_i$, we again consider spatial
bins indexed by $j$, allowing us to estimate the actual (initial)
velocity dispersions of our particular realization. For example, the
average radial dispersion in bin $j$ is given by
\begin{equation}
\left[\sigma_R^2\right]_j^{\rm actual} = \frac{1}{M_j} \sum_{\vec{\hat
    x}_i \, {\rm in \, cell} \, j} m_p (\vec{\hat v}_i \cdot
\vec{e}_R^{(i)})^2 .
\end{equation}
The normalization factor
\begin{equation}
M_j =  \sum_{\vec{\hat x}_i \, {\rm in \, cell} \, j} m_p 
\end{equation}
is simply equal to the mass of the initial realization that falls into
the spatial bin. The vector $\vec{e}_R^{(i)}$ is the radial unit
vector at the position of particle $i$. The expected target velocity
dispersion in the bin is given by
\begin{equation}
\left[\sigma_R^2\right]_j^{\rm target} = \frac{1}{M_j^0} \int_{{\rm
    cell} \, j} \rho_0(\vec{x}) \, \sigma^2(\vec{x}) \, {\rm
  d}\vec{x},
\end{equation}
where ${M_j^0}$ is the mass expected in the continuum in the cell.  We
may then define a merit function that measures the deviation of the
actually realized velocity dispersion relative to the target value. To
this end we adopt
\begin{equation}
Q_R = \sum_j \frac{|\left[\sigma_R^2\right]_j^{\rm actual} -
  \left[\sigma_R^2\right]_j^{\rm target}|}{
  \left[\sigma_R^2\right]_j^{\rm target}}.
\end{equation}
Similarly, we define merit functions $Q_z$ and $Q_\phi$ for the
dispersions in the $z$-direction and the azimuthal $\phi$-direction,
respectively. In the azimuthal direction, we actually measure and
constrain both the full second moment $\left<v_\phi^2\right>$ and the
dispersion relative to the mean streaming $\sigma_\phi^2$. Combined,
this then also constrains the mean streaming itself.

As for the density case, we need to adopt a suitable discretization
for the spatial bins. In order to avoid biases due to different noise
levels in the bins, we adopt bins in which the mass per bin (or,
equivalently, the number of particles) is roughly constant, a
situation we realize with the help of a hierarchical adaptive binning
scheme. We note that it is possible to employ the same spatial bins as
used for the density merit function, but if desired the corresponding
target value for the mass per bin can also be chosen differently.

\subsection{Optimization procedure}

Our goal is to iteratively adjust the $\vec{\hat v}_i$ such that $S$,
$Q_R$, $Q_\phi$ and $Q_z$ are simultaneously minimized. We do this by
combining these quantities into a single goodness-of-fit parameter,
\begin{equation}
S_{\rm global}  = S + \chi (Q_R + Q_\phi + Q_z),
\end{equation}
where the constant $\chi$ is adjusted such that $S$ and $Q_R + Q_\phi
+ Q_z$ are of the same magnitude and have the same units. In other
words, we give equal weight to the density and velocity constraints.

The function $S_{\rm global}(\vec{\hat v}_1, \ldots, \vec{\hat v}_N)$
depends only on the initial particle velocities. Hence we are formally
charged with the task to find its minimum in the high-dimensional
space of all the $3 N$ velocity components. Trying to find this
minimum is a computationally rather tricky problem, because the
function will feature a large number of local minima in which a direct
search may easily get stuck. Also, the function is non-linear and
expensive to evaluate -- calculating $S$ involves orbit integrations
of a large number of particles over a long time interval. Even if a
single force calculation is comparatively cheap due to the static
potential, the cumulative CPU cost can become demanding, especially
since we are not dealing just with a single particle but rather with a
(potentially quite large) particle collection of size $N$.

Nevertheless, it is still possible to estimate the local gradient of
$S_{\rm global}$ with respect to the $\vec{\hat v}_i$ and then to move
in the direction of steepest decent by simultaneously modifying all
velocities in the direction opposite to the gradient. But finding a
local minimum in this way will still be very hard (we have tried);
typically, one will instead overshoot in at least one of the many
dimensions of the problem.

Another consideration also argues against this brute force
approach. Physically, we expect that the particles should be
completely uncorrelated in proper collisionless initial
conditions. Directly minimizing $S_{\rm global}$ simultaneously with
respect to all velocities invokes the danger of `overfitting', where a
low value of the merit function is obtained through the introduction
of velocity correlations in the specific N-body realization of the
system.

Our solution to these problems involves two components. First, we
serialize the minimization procedure, i.e.~we always pick only one
particle randomly, and then optimize its velocity such that $S_{\rm
  global}$ is reduced. Second, we do not actually try to adjust the
velocity of the single particle such that $S_{\rm global}$ is
necessarily minimized, as may be done by using the result of a line
search along a single parameter. Rather, we simply randomly pick a new
guess for the particle's velocity and (re)evaluate the merit function
for this choice. If the proposed velocity improves the fit, we retain
it as the new velocity of the particle, otherwise we simply keep the
particle's old velocity and proceed with the next particle. This is
simply repeated until the fit cannot be improved significantly any
more.  We note that this approach bears some resemblance to Monte
Carlo Markov Chain techniques, except that we are here trying to find
a global optimum rather than exploring a likelihood surface where one
also moves occasionally away from the optimum with a certain
probability.

The distribution from which one draws the trial velocities is in
principle arbitrary, provided it is broad enough to sample all allowed
velocities. However, it is highly advantageous to make it close to the
target distribution function, because in this case the convergence
speed can be expected to be particularly rapid (just as in MCMC). In
our case, we can simply use Gaussians for that, as we already have the
second moments in hand based on our Jeans solutions and the
distribution function will in most cases resemble a Gaussian locally,
so this should facilitate rapid convergence.  Note that every new
trial velocity we pick is completely independent of the previous
value, as well as of the velocities of all other particles. This helps
to minimize correlations between different particles in the created
initial conditions, and it prevents to get easily stuck in a local
minimum. Nevertheless, velocity correlations are not
  completely absent, because the acceptance decision for the velocity
  of a particular particle still depends on the discrete spectrum of
  velocities realized at this instant for all the other particles. But
  as our results show, any present residual correlations do not seem
  to negatively impact the quality of the created initial conditions.

In practice, we choose to process all particles in a random order. In
each pass over the particles, we pick for a given particle one of its
three principal coordinate directions and draw a random trial value
for the corresponding velocity component.  We note that the evaluation
of $S_{\rm global}$ can be significantly accelerated if only one
particle is varied. In this case, only the summed orbital response of
all particles needs to stored, without requiring storage of all the
responses individually. Evaluating $S_{\rm global}$ for a changed
velocity of one particle then boils down to calculating the orbit
response for this particle twice, both for the old and new
velocities. The differential between the two results can then be
appropriately added to the global response to assess the change in
$S_{\rm global}$.

\section{Velocity constraints}  \label{sec:velconstraints}

As discussed above, a problematic aspect of optimizing only a density
merit function is that it is ambiguous to which solution this will
converge. Recall that for a given density distribution there will in
general be a vast number of possible distribution functions.  The
iterative optimization will yield a particular realization of one of
these distribution functions, and this solution might depend on the
initial velocity guesses one has used at the beginning.  In order to
make the solution well defined, we need to impose additional
constraints that reflect the desired properties of the specific
solution one is looking for. We do this in terms of second moments of
the velocity distribution and by forcing the system to converge to a
solution that features these moments. The moments themselves are
calculated from the Jeans equations. Different possibilities for a
specification of the desired properties of the target system exist.

\subsection{Spherically symmetric distribution functions}

If the density structure is spherically symmetric and the velocity
distribution function depends at most on the magnitude of the angular
momentum, we can make use of the spherically symmetric Jeans equation
for the second radial velocity moment,
\begin{equation}
\frac{\partial(\rho\sigma_r^2)}{\partial r} + 2 \frac{\beta \rho
  \sigma_r^2}{r} + \rho\frac{\partial \Phi}{\partial r} =
0. \label{eqn:jeans_spherical}
\end{equation}
Here $\sigma_r^2 = \left< v_r^2 \right>$ is the radial dispersion.
The velocity distribution functions in the transverse directions at
any given position need not be equal to that in the radial direction,
but we have $\sigma_\theta = \sigma_\phi$ due to the assumed symmetry.
The degree of radial--tangential anisotropy is usually measured in
terms of
\begin{equation}
\beta = 1 - \frac{\sigma_t^2}{2 \sigma_r^2} = 1 -
\frac{\sigma_\phi^2}{\sigma_r^2} ,
\end{equation}
where $\sigma_t^2 = \sigma_\theta^2 + \sigma_\phi^2 = 2
\sigma_\theta^2$ measures the total tangential dispersion, and due to
spherical symmetry we have $\sigma_\theta = \sigma_\phi$.  If the
distribution is isotropic, we have $\beta=0$. If the orbits are biased
towards radial motions we have $\beta > 0$, while for $\beta <0$ they
are preferentially tangential.

For given $\rho(r)$ and prescribed $\beta(r)$, and thanks to the
purely radial dependence, equation~(\ref{eqn:jeans_spherical}) becomes
an ordinary differential equation for $\rho \sigma_r^2$ which can be
readily integrated using the boundary condition $\rho \sigma_r^2=0$
for large radii. Dividing the solution by the density then yields the
dispersion $\sigma^2_r(r)$ as a function of radius, and from it we
also obtain $\sigma^2_\theta(r) = (1-\beta) \sigma^2_r(r)$.

We note that we may choose $\beta$ to be a function of radius, as
suggested by the structure measured for cosmological dark matter
halos. \citet{Hansen2006} found that the local anisotropy of dark
matter halos correlates well with the logarithmic slope
\begin{equation}
\alpha = \frac{ d  \ln \rho}{{d}\ln r}
\end{equation}
of the density profile. Their numerical results are well
fit by the relation
\begin{equation}
\beta(r) = -0.15 - 0.2 \alpha , 
\end{equation}
which we adopt as an additional option in our IC code. This implies
nearly isotropic orbits in the center of a Hernquist or NFW halo, and
a growing preference for more radial dispersion as a function of
distance.  
 
A particularly simple choice for $\beta(r)$ is the isotropic case,
$\beta=0$, where the velocity distribution function is independent of
direction at every point. In this ergodic case, the distribution
function depends only on energy.  \citet{Hernquist1990} constructed
such a solution for a density profile of the form $\rho(r) \propto
r^{-1} (r+a)^{-3}$, which is cosmologically particularly relevant as
it has a shape similar to the NFW density profile \citep{Navarro1997}
measured for relaxed halos in cold dark matter structure formation
simulations. This makes the isotropic Hernquist model a particularly
useful analytic distribution, and we will also use it here to verify
our procedures. We note however that a yet more realistic model would
be one with a radially varying anisotropy $\beta(r)$. No analytic
distribution functions are known for this case, but such models can be
readily constructed with our new method.

\begin{figure}
\begin{center}
\resizebox{8cm}{!}{\includegraphics{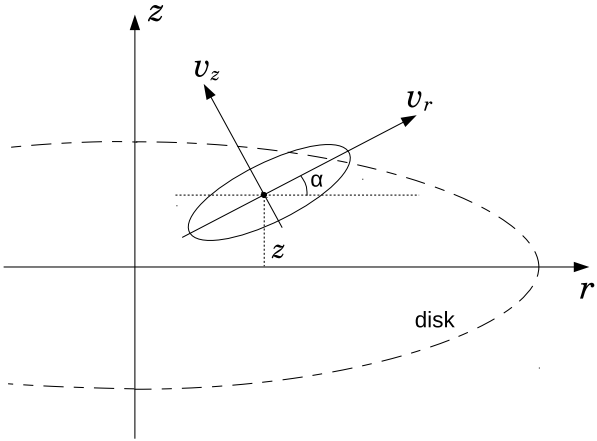}}
\caption{Sketch of the geometry adopted to describe the assumed tilt
of the velocity ellipsoid in the $f(E, L_z, I_3)$ case.}
\label{fig:sketch}
\end{center}
\end{figure}

\subsection{Axisymmetric systems with two integrals of motion}

For axisymmetric systems, the angular momentum $L_z$ around the
$z$-axis is a conserved quantity for all orbits, hence we expect the
distribution function to depend on $L_z$ besides energy $E$. In
general, there can be a third integral of motion, $I_3$, which is
however often not easy to identify and therefore considered
``non-classical''. If one disregards $I_3$ and assumes that the
distribution function is only a function of $E$ and $L_z$, then the
situation simplifies considerably, as one can then infer that all
mixed moments of the velocity distribution vanish (i.e. $\left<
\sigma_R \sigma_z\right>=0$). In this case the axisymmetric Jeans
equations simplify considerably and can be comparatively easily
solved.

With two integrals of motion, the non-trivial axisymmetric Jeans
equations become:
\begin{equation}
\frac{\partial(\rho\sigma_z^2)}{\partial z} + \rho\frac{\partial
  \Phi}{\partial z} = 0, \label{eqn:jeans_axi1}
\end{equation}
and
\begin{equation}
\left< v_\phi^2 \right> = \sigma_R^2 +\frac{R}{\rho} \frac{\partial (
  \rho\sigma_R^2)}{\partial R}+ R \frac{\partial \Phi}{\partial R}.
 \label{eqn:jeans_axi2}
\end{equation}
The mean streaming motions in the radial and vertical directions
vanish, $\left<v_R\right> = \left<v_z\right> =0$ (but not necessarily
in the azimuthal direction), and importantly, the radial and vertical
dispersions are equal everywhere, $\sigma_R^2 = \sigma_z^2$.

This in particular means that the density distribution fully specifies
the vertical and radial dispersions in the meridional plane $(R,z)$. 
They can be explicitly
calculated as
\begin{equation}
\sigma_R^2 = \sigma_z^2(R,z) = \frac{1}{\rho(R,z)} \int_z^\infty
\rho(z',R) \frac{\partial \Phi}{\partial z}(R, z') \,{\rm d}z'.
\end{equation}

Once these dispersions are known, we can now determine the second
moment $\left< v_\phi^2 \right>$ of the azimuthal motion from the
radial Jeans equation (\ref{eqn:jeans_axi2}).  However the mean
streaming $\left<v_\phi\right>$ in the azimuthal direction is not
specified by the Jeans equations. Indeed, $\left<v_\phi\right>$ does
not have to be zero if there is net rotation. For any given solution
with non-zero $\left<v_\phi\right>$, one can readily construct new,
equally valid equilibrium solutions, for example by reversing all or a
fraction of the particles' $\phi$-motions. It is hence clear that the
requirement of axisymmetry does not specify $\left<v_\phi\right>$. In
fact, we are (within limits) free to set this.

We adopt the parameterization 
\begin{equation}
\left<v_\phi\right>^2 = k^2 \left[ \left< v_\phi^2 \right> -
  \sigma_R^2 \right]
\label{eqnsatoh}
\end{equation}
suggested by \citet{Satoh1980} to specify the mean streaming.
For the interesting choice $k=1$, we obtain for the azimuthal dispersion
\begin{equation}
\sigma_\phi^2 \equiv \left< v_\phi^2\right> - \left<v_\phi \right>^2 =
\sigma_R^2 = \sigma_z^2 ,
\label{eqn:streaming}
\end{equation}
i.e.~$\sigma_\phi^2$ is then equal to the radial and vertical
dispersions. This defines the case of an isotropic rotator.  But we
may also adopt a lower or higher value for $k$, or even one with a
spatial dependence, up to the maximum allowed local value of
\begin{equation}
k_{\rm max}^2 =
\frac{\left<v_\phi^2\right>}{ \left<v_\phi^2\right> - \sigma_R^2}.
\end{equation}
In $k$ climbs up to this value, the azimuthal dispersion vanishes and
we have $\left<v_\phi\right>^2 = \left<v_\phi^2\right>$, corresponding
to a system with the maximum possible angular momentum for a given
density structure.  In our {\small GALIC} code, we either choose a
constant $k$ or specify $k$ in units of $k_{\rm max}$ when the case of
a $f(E, L_z)$ distribution function is selected.

\begin{figure*}
\begin{center}
\setlength{\unitlength}{1cm}
\begin{picture}(15.0,10)

\put(0,5.1){\resizebox{5cm}{!}{\includegraphics{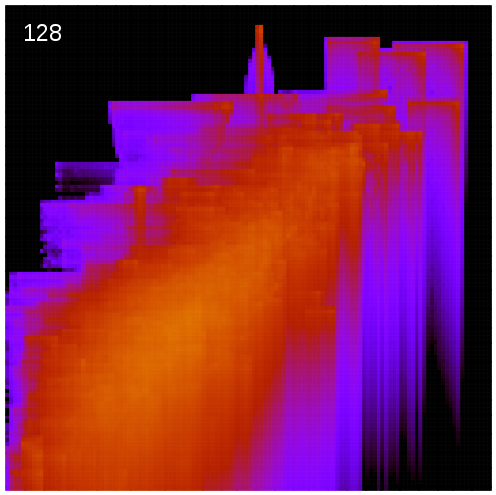}} }
\put(5.1,5.1){\resizebox{5cm}{!}{\includegraphics{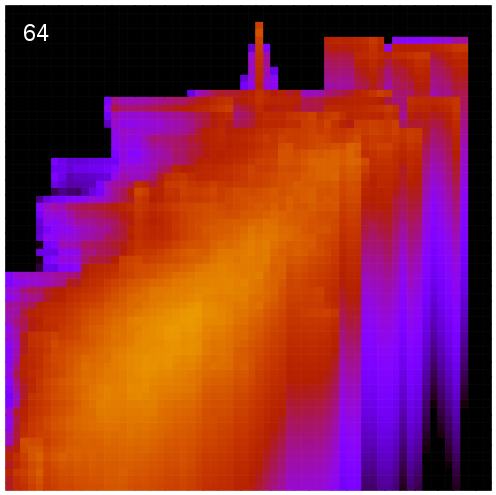}} }
\put(10.2,5.1){\resizebox{5cm}{!}{\includegraphics{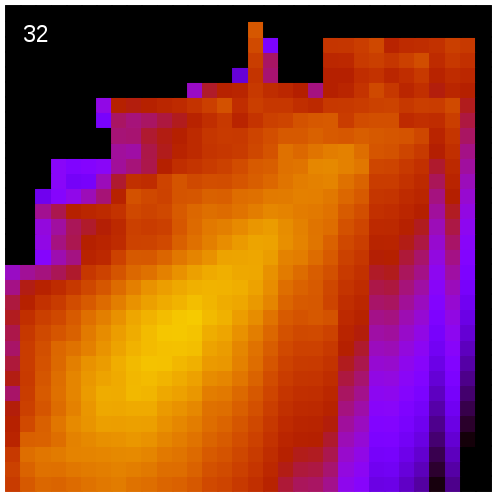}} }
\put(0,0){\resizebox{5cm}{!}{\includegraphics{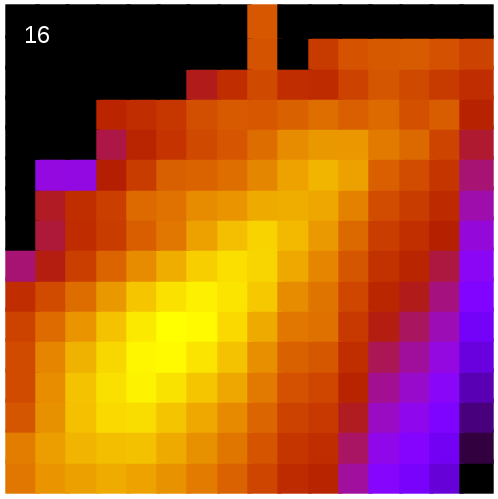}} }
\put(5.1,0){\resizebox{5cm}{!}{\includegraphics{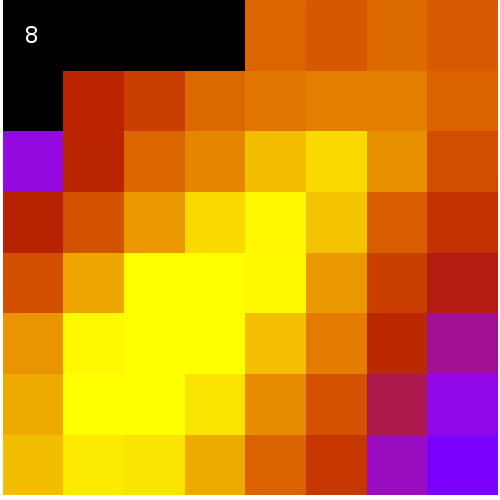}} }
\put(10.2,0){\resizebox{5cm}{!}{\includegraphics{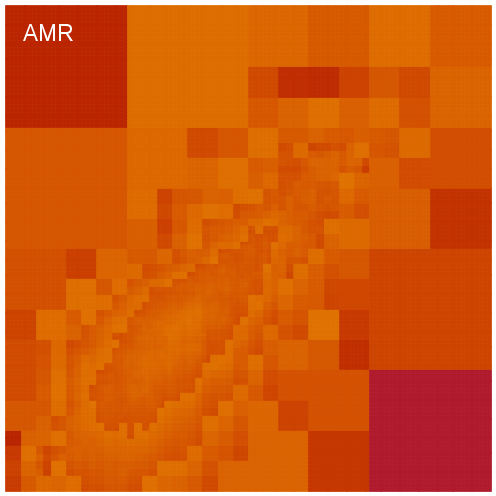}} }
\end{picture}
\caption{Density response of our hierarchical binning scheme. The top
  left panel shows the average orbit response as recorded on the
  finest grid used (which is a logarithmic grid with $256^2$ pixels).
  The next four panels show coarsened representations of this field,
  with resolutions of $64^2$ to $8^2$ pixels (this continues to even
  coarser meshes that are not shown). Finally, the bottom right panel
  shows the variable resolution response that is actually used to
  compare with the target density distribution, based on bins containing
  roughly equal mass.
\label{fig:densreponse}
}
\end{center}
\end{figure*}

\subsection{General systems with three integrals of motion}

While simple disk models can be constructed as isotropic rotators,
observations in the Milky Way at the Solar circle suggest that
$\sigma_R$ is not equal to $\sigma_z$. Rather, the two dispersions are
related approximately by $\sigma_z \simeq 0.5 \, \sigma_R$
\citep{Binney1998}. Even if the Milky Way can still be described well
as an axisymmetric system, this already means that the distribution
function is not only dependent on $(E,L_z)$; instead, a third integral
of motion must play an important role. In the outer parts of the disk,
this is approximately given by the energy of the vertical motion in
the disks potential, but in the inner parts of the disk this
identification presumably becomes a poor approximation.

\begin{table*}
\begin{tabular}{lp{7cm}p{9cm}}
\hline
Model & Components and their shape & Imposed velocity structure   \\
\hline
 H1   &  spherical dark halo  &  ergodic (i.e. isotropic Hernquist model)\\
 H2   &  spherical dark halo  &  spherical anisotropy, radial orbits dominating, $\beta = 0.5$ \\
 H3   &  spherical dark halo  &  spherical anisotropy, tangential orbits dominating, $\beta = -1.0$ \\
 H4   &  spherical dark halo  &  radially varying anisotropy, $\beta(r) = -0.15 -0.20 \frac{d\log\rho}{d\log r}$\\
 H5   &  spherical dark halo  &  axisymmetric velocity structure, isotropic rotator with $k=1$ \\
 H6   &  prolate dark halo with $s=0.85$ &  axisymmetric velocity structure, no net rotation \\
 H7   &  oblate dark halo with $s=1.15$ &  axisymmetric velocity structure, no net rotation \\
\hline
 B1   &  spherical dark halo, spherical bulge & ergodic \\
 B2   &  spherical dark halo, spherical bulge & different anisotropies for bulge and halo, $\beta_{\rm halo} = 0.5$, 
$\beta_{\rm bulge} = -1.0$ \\
 B3   &  prolate dark halo $s=0.85$, spherical bulge & axisymmetric velocity structure, no net rotation \\
 B4   &  oblate dark halo $s=1.15$, prolate bulge  $s=0.85$ & axisymmetric velocity structure, no net rotation \\
\hline
 D1   & spherical dark halo, thin disk & axisymmetric velocity structure for halo and disk, disk isotropic rotator\\ 
 D2   & prolate dark halo with $s=0.85$, thin disk & axisymmetric velocity structure for halo and disk, disk isotropic rotator\\ 
 D3   & spherical dark halo, thin disk &  disk with $f(E,L_z,I_3)$ structure
and $f_{R}=2.0$, halo axisymmetric with $k=0$ \\ 
 D4   & spherical dark halo, thin disk & disk with $f(E,L_z,I_3)$ and $f_{R,{\rm disk}}=4.0$, halo axisymmetric with $k=0.5$   \\
 D5   & prolate dark halo with $s=0.85$, thin disk & disk with $f(E,L_z,I_3)$ 
and $f_{R,{\rm disk}}=2.0$, halo axisymmetric isotropic rotator\\ 
\hline
 M1  & spherical dark halo, spherical bulge, thin disk & axisymmetric structure for halo and bulge (no rotation), disk isotropic rotator\\ 
 M2  & spherical dark halo, spherical bulge, thin disk & axisymmetric velocities for halo/bulge,  disk with $f(E,L_z,I_3)$, $f_{R,{\rm disk}}=2.0$\\ 
 M3  & spherical dark halo, spherical bulge, thin disk & disk with $f(E,L_z,I_3)$ and $f_{R}=4.0$, bulge no rotation, halo with $k=0.1$\\ 
 M4  & prolate dark halo $s=0.85$, oblate bulge $s=1.15$, thin disk & disk with $f(E,L_z,I_3)$, $f_{R}=2.0$, halo and bulge both isotropic rotators\\ 
\hline
\end{tabular}
\caption{Set of basic galaxy models constructed for testing purposes
  with the methods outlined in this paper. Unless stated otherwise, we
  have used $10^6$ particles for each model component. The models
  labeled `H1', `H2', etc., contain only a dark matter halo but differ
  in the halo shape or the assumptions made for the velocity
  structure. The models denoted `B1', `B2', and so on, contain a bulge
  in addition to the halo, and the modes with `D1', `D2', etc.,
  feature a disk in addition to the halo. Finally the 
  models `M1', `M2', etc., contain both a stellar bulge and a stellar
  disk, next to a dark matter halo.
  \label{TabSims}}
\end{table*}

Another interesting observational fact is that the velocity ellipsoid
above the disk mid plane is not aligned with the coordinate plane;
instead, it appears tilted (i.e.~$\left<v_Rv_z\right> \ne 0$) towards
the centre of the system. Using RAVE velocity data, \citet{Siebert2008} and
\citet{Binney2014} quantified the tilt at $1\,{\rm kpc}$ above the
disk to be around $\alpha = 7^o$ (see sketch of
Fig.~\ref{fig:sketch}). We are hence forced to apply the general
axisymmetric Jeans equations, which take the form:
\begin{equation}
\frac{\partial(\rho\sigma_z^2)}{\partial z} + \rho\frac{\partial
  \Phi}{\partial z} + \frac{1}{R} \frac {\partial (R\rho
  \left<v_Rv_z\right>)}{\partial R} = 0, \label{eqn:jeans_axi3}
\end{equation}
\begin{equation}
\left< v_\phi^2 \right> = \sigma_R^2 + \frac{R}{\rho} \frac{\partial (
  \rho\sigma_R^2)}{\partial R} + R \frac{\partial \Phi}{\partial R} +
\frac{R}{\rho} \frac{\partial ( \rho \left<v_Rv_z\right> )}{\partial
  R},
 \label{eqn:jeans_axi4}
\end{equation}
\begin{equation}
\frac{\partial (\rho \left<v_z v_\phi\right>)}{\partial z} +
\frac{1}{R^2}\frac{\partial(R^2 \rho \left< v_R v_\phi
  \right>)}{\partial R} = 0 .
 \label{eqn:jeans_axi5}
\end{equation}
This system of equations is significantly under-specified and
additional assumptions are needed for closure. We shall assume that
the velocity ellipsoid is not tilted in the $\phi$-direction, hence
$\left<v_z v_\phi\right> = \left< v_R v_\phi \right> = 0$. This
eliminates the third equation. However, we need to retain a tilt in
the meriodonal plane, as encoded by $\left<v_Rv_z\right>$. If $\alpha$
is the local angle between the velocity ellipsoid and the $R$-axis,
this mixed moment can be expressed in terms of the radial and vertical
moments, i.e. we have
\begin{equation}
\left<v_Rv_z\right> = \frac{1}{2} \tan(2\alpha)\left[ \sigma_R^2 -
  \sigma_z^2 \right].
\end{equation}
For reference, the dispersions in the rotated coordinate frame
$(R',z')$ are given by
\begin{equation}
\left< v_{R'}^2\right> = \sigma_R^2 \cos^2(\alpha) +
\left<v_Rv_z\right> \sin(2\alpha) + \sigma_z^2 \sin^2(\alpha),
\end{equation}
\begin{equation}
\left< v_{z'}^2\right> = \sigma_R^2 \sin^2(\alpha) -
\left<v_Rv_z\right> \sin(2\alpha) + \sigma_z^2 \cos^2(\alpha).
\end{equation}
The tilt angle is the one for which $\left <v_{R'} v_{z'}\right>=0$,
by construction.

Interestingly, the tilt observed for the Galaxy at the Solar circle is
consistent with the velocity ellipsoid pointing approximately to the
center of the Galaxy; the most recent determination by
\citet{Binney2014} gives $\alpha \sim 0.8\, \arctan(z/R)$. We here
assume for definiteness that this alignment is perfect and holds
throughout the system, in which case the angle $\alpha$ is simply
given by
\begin{equation}
\tan \alpha = \frac{z}{R}.
\end{equation}
Specifying the orientation of the velocity ellipsoid in this way has
the nice property of naturally producing a spherically symmetric
orientation close to the galactic centre, i.e. the `disk regime'
seamlessly transitions to a `bulge regime'.  Far out in a thin disk,
the velocity ellipsoid will align with the coordinate axes, while near
to the centre the situation becomes closer to that in a spherically
symmetric case with a radial alignment, which seems plausible.

Prescribing the tilt angle is not yet enough to solve equations
(\ref{eqn:jeans_axi3}) and (\ref{eqn:jeans_axi4}), because they
involve four unknowns. An additional assumption is required. To this
end, we adopt a prescribed relation between the radial and vertical
dispersions in the tilted velocity ellipsoids, namely
\begin{equation}
\left< v_{R'}^2\right> = f_R \left< v_{z'}^2\right>,
\end{equation}
where $f_R$ is a factor specifying the anisotropy between radial and
transverse motions.  For the disk of a Milky Way like galaxy, we would
expect $f_R \simeq 2$ at the Solar circle, but little is own about a
potential radial variation of this value.  We also note in passing
that the Toomre stability criterion depends sensitively on $\sigma_R$,
so invoking values $f_R > 1$ is one way of stabilizing a stellar disk
of given thickness against axisymmetric perturbations.  For
simplicity, we shall assume a spatially constant value for $f_R$ in
the disk, but note that our techniques could be easily generalized to
include a radial or vertical variation of this factor.

Given the above model for the dispersions, we can now express the
mixed moment $\left<v_Rv_z\right>$ through the vertical dispersion,
namely
\begin{equation}
\left<v_Rv_z\right> = h\, \sigma_z^2,
\end{equation}
where the function $h = h(R,z)$ is given by
\begin{equation}
h= \frac{(f-1)\tan(2\alpha)}{2\cos^2(\alpha) - 2f \sin^2(\alpha) +
  (1+f) \sin(2\alpha)\tan(2\alpha)},
\end{equation}
and the shortcut $f=f_R$ is understood.  The Jeans equation
(\ref{eqn:jeans_axi3}) now becomes an inhomogenous first order partial
differential equation (PDE) for $\sigma_z^2$. Defining $q \equiv \rho
\sigma_z^2$, the relevant equation takes the form
\begin{equation}
\frac{\partial q}{\partial z} 
+\frac{\partial(h q)}{\partial R} + \frac{hq}{R} 
+
 \rho\frac{\partial
  \Phi}{\partial z} 
 = 0. \label{eqn:jeans_pde}
\end{equation}
We can solve this PDE numerically with the methods of lines by
discretizing in $R$ and replacing the spatial $R$-derivative with a
finite difference approximation. We can then integrate the resulting
system of coupled ordinary differential equations along the
$z$-direction, starting at $z\simeq \infty$ and ending up at
$z=0$. The initial condition is $q(R, z=\infty) = 0$, augmented with
the boundary condition $q(R=\infty, z) = 0$. For numerical stability,
one needs to take care that an upwind finite difference estimate for
the $R$-derivative is used. Note also that the $hq/R$ term is not
singular for $R=0$, because $h/R \to (f-1)/f$ for $R\to 0$.

Having obtained a solution for $q(R,z)$, we then readily have
$\sigma_z$, $\sigma_R^2$, $\left<v_Rv_z\right>$, as well as
$\left<v_{R'}^2\right>$ and $\left<v_{z'}^2\right>$ throughout the
meridional plane. Similar as with the axisymmetric $f(E,L_z)$ case, we
still have the freedom to choose a streaming velocity in the
$\phi$-direction, except that now we have to use equation
(\ref{eqn:jeans_axi4}) to infer the corresponding dispersion available in
the azimuthal direction. We continue to use the parametrization of
equation (\ref{eqn:streaming}) for the azimuthal streaming. For the
case $k=1$, we then get $\sigma_\phi^2 = \sigma_R^2$ in the mid-plane.

We note that one can also obtain from epicycle theory a statement
about the relation between $\left< (v_\phi - v_c)^2\right>$ and
$\sigma_R^2$, valid for small radial dispersions $\sigma_R$, namely
\begin{equation}
\frac{\left< (v_\phi - v_c)^2\right>}{\sigma_R^2} \simeq
\frac{1}{\gamma^2}
\label{eqnepicycle}
\end{equation}
where $\gamma= 2 \Omega / \kappa$. $\Omega^2 =
\frac{1}{R}\frac{\partial \Phi}{\partial R}$ is the circular orbit
frequency, and
\begin{equation}
\kappa^2 = R\frac{{\rm d}\Omega^2}{{\rm d}R} + 4 \Omega^2
\end{equation}
is the epicycle frequency. Typically we have $1/\gamma^2 \simeq 0.5$.
We note that equation (\ref{eqnepicycle}) is only reliable for very
cold thin disks, with $\sigma_R \ll v_c$
\citep[see][]{BinneyTremaine2008}. Interestingly, combined with
equation (\ref{eqn:jeans_axi4}), the epicycle approximation gives the
azimuthal streaming (and hence also the axisymmetric drift) in the
equatorial plane as
\begin{equation}
\left<v_\phi\right> = v_c + \frac{\sigma_R^2}{2v_c}\left(
\frac{\partial \ln( \rho \sigma_R^2)}{\partial
  \ln R} +  \frac{\gamma^2  - 1}{\gamma^2}\right).
\end{equation}
On the other hand, we obtain from equation (\ref{eqnsatoh}) the
following expression for the streaming velocity to leading order in
$\sigma_R / v_c$:
\begin{equation}
\left<v_\phi\right> = k\, v_c + \frac{k\, \sigma_R^2}{2v_c}\left(
\frac{\partial \ln( \rho \sigma_R^2)}{\partial
  \ln R} \right).
\end{equation}
Consistency with the epicycle approximation hence requires $k=1$ for
thin cold disks. The residual difference grows for large $\sigma_R /
v_c$, but note that in this limit the epicycle approximation becomes
inaccurate anyway.

\section{Implementation details}  \label{sec:implementation}

\subsection{Adaptive logarithmic binning}  \label{secbinning}

To account for the typical power-law growth of the density towards the
center in self-gravitating systems, we generally employ logarithmic
grids.  For the sake of simplicity, we restrict ourselves to
axisymmetric systems in this paper, and also assume mirror symmetry
with respect to the $z=0$ plane. Adopting cylindrical coordinates,
this means we only have to cover the positive quadrant in the
$(R,z)$-plane. We assume that the mass distribution is fully contained
inside a cube of side-length $2d_{\rm max}$, i.e. our mesh needs to
cover the region $0 \le R < d_{\rm max}$ and $0 \le z < d_{\rm
  max}$. If we use $N_{\rm bin} = 2^{l}$ bins per dimension, and
require that the width of the bins grows by a constant factor $f$ from
bin to bin, the borders of the bins can be written as
\begin{equation}
R_i  =  d_{\rm base} (f^i - 1),   
\end{equation} 
\begin{equation}
z_j  =  d_{\rm base} (f^j - 1),
\end{equation} 
with $i,j \in [0, 1, \ldots, N_{\rm bin}]$. The bin $({i,j})$ (with
$0\le i,j < N_{\rm bin}$) then covers $[R_i, R_{i+1}] \times [z_j,
  z_{j+1}]$ in the $(R,z)$-plane and has volume
\begin{equation}
V_{ij} =  
2\pi (R_{i+1}^2-R_i^2)(z_{j+1} - z_j).
\end{equation}
To cover the full volume, $d_{\rm base}$ and $f$ need to be chosen
such that
\begin{equation}
d_{\rm max}  =  d_{\rm base} (f^{N_{\rm bin}} - 1).   
\end{equation} 
This still leaves room for one additional constraint to fully specify
the quantities $d_{\rm base}$ and $f$. We typically address this by
requiring that the first bin, bounded by $R_1 = d_{\rm base} (f - 1)$,
encloses a small prescribed fraction of the total mass of the system
(e.g. $10^{-6}$), such that the central region is still well resolved
by the grid.

As we discussed earlier, the objective functions assessing the density
response and the initial velocity distribution work best if the
spatial bins are chosen such that they contain approximately constant
mass. We realize such a scheme by first constructing the mass response
on a relatively fine grid, given by the level $l_{\rm max}$. We then
recursively construct a set of coarsened meshes on levels $l_{\rm max}
-1$, $l_{\rm max} -2$, $\ldots$, $1$, $0$, until there is only one
cell left covering the whole quadrant. Computing the mass response of
one of the grid cells of a coarsened mesh is done recursively by
summing over the corresponding four cells in the finer mesh one level
higher. Similarly for the velocity dispersion fields.

Evaluating the objective functions then proceeds with a recursive
algorithm that walks the tree of nested mesh cells. Beginning at the
`root node', a mesh cell is included in the sum if it contains less
than a certain threshold mass or if it is already a cell of the finest
level. Otherwise, the mesh cell is `opened', and its four daughter
cells are considered in turn as candidates for being included in the
sum. This procedure automatically selects a close to optimum set of
cells of different sizes. Note that the union of the cells that enter
the sum form a space-covering tessellation, i.e. each point in the
$(R,z)$-plane is accounted for exactly once. 

In Figure~\ref{fig:densreponse} we show an example for the mass
response grid of a set of orbits in the $(R,z)$-plane for a mesh with
$N_{\rm bin} = 2^8$ cells per dimension, together with the hierarchy
of the next four coarsened representations at higher levels. The final
panel on the bottom right shows a mixed image of variable resolution,
indicating what is effectively used in the adaptively calculated sum
that defines the merit function.

\begin{figure}
\begin{center}
\resizebox{8.5cm}{!}{\includegraphics{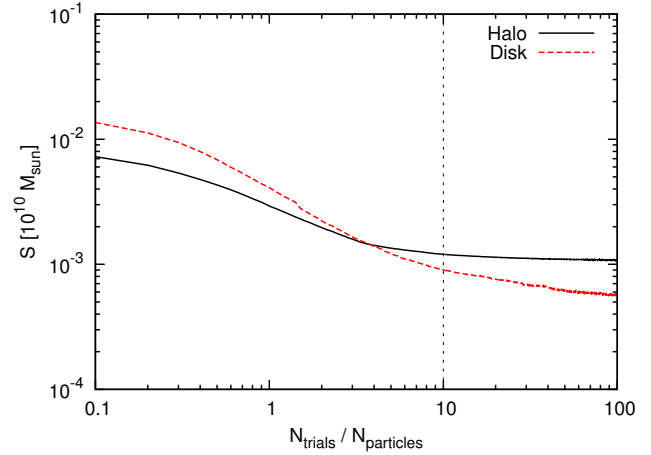}}
\caption{Decline of the merit function in a typical optimization run
  (actually the model D1 from Table~1) as a function of the number of
  attempted velocity adjustments in units of the particle number of
  the corresponding component.  The solid line shows the result for
  the dark matter particles, while the dashed lines is for the disk
  particles.}
\label{fig:convergence}
\end{center}
\end{figure}

\begin{figure*}
\includegraphics[natwidth=1bp,natheight=1bp, width=7.2cm,
  angle=-90]{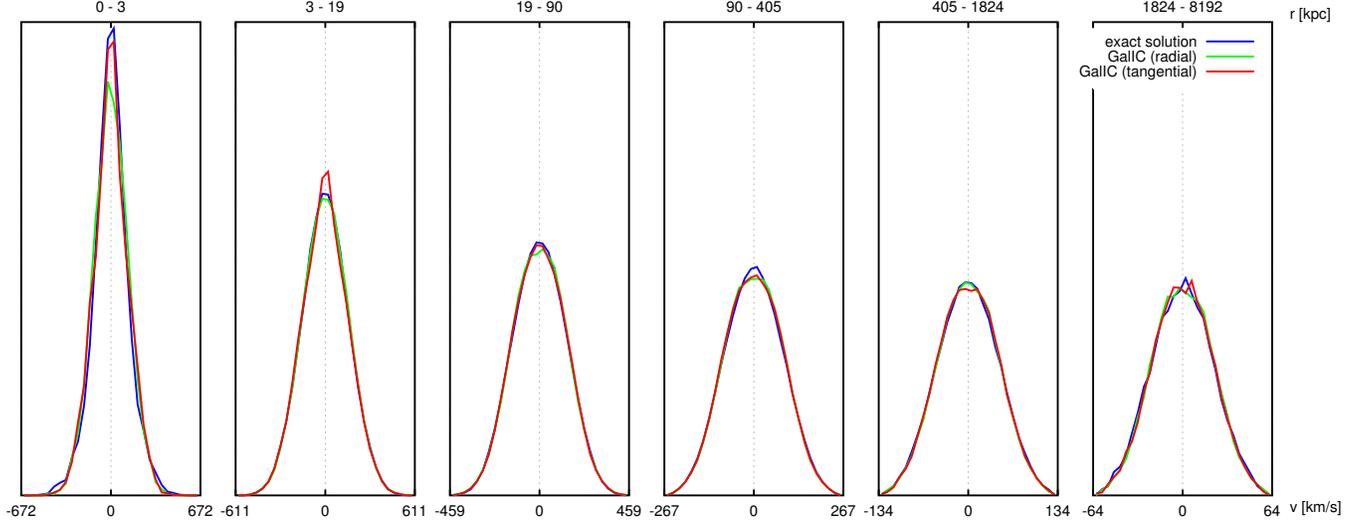}
  \caption{Radial and  azimuthal velocity distribution functions in
    different radial shells for the isotropic Hernquist sphere. The
    blue line show the exact analytic solution. The red and green
    lines show the constructed solution with our method, separately
    for  azimuthal and radial velocity components. The numbers at the
    top of each panel indicate the radial range of the measurement. }
\label{fig:HqVelRadial}
\end{figure*}

\subsection{Orbit integration}

In order to efficiently and accurately compute orbits of particles for
arbitrary mass distributions, we produce a look-up table of the
gravitational potential and its derivatives $\partial \Phi / \partial
R$ and $\partial \Phi / \partial z$ in the positive quadrant of the
$(R,z)$-plane.  Due to the axisymmetry we assume, this is sufficient
to obtain the forces and the potential everywhere through a table
look-up.  We use a fine logarithmic grid in $R$ and $z$ and bilinear
interpolation for the look-up table.

In order to allow a computation of the forces for arbitrary density
distributions without analytic solutions of Poisson's equation, we
numerically evaluate the potential and forces on the fine grid based
on randomly sampling the density distribution with a very large number
of fiducial particles combined with a calculation of the forces and
potentials with a gravitational tree algorithm. This procedure is very
flexible and accurate. In order to help reducing noise effects from
the sampling to a negligible level, a large number of fiducial points
is used, and additionally, we evaluate and average the field at a set
of different azimuthal angles.

For the orbit integration of particles, which gives us $\vec{x}_{\rm
  orbit}(\vec{\hat x}_i, \vec{\hat v}_i, t)$, we use the leapfrog
scheme with adaptive timestep based on the kick-drift-kick
formulation. If $(\vec{x}^{(n)}, \vec{v}^{(n)})$ denote position and
velocity after step $n$, then the update to the next step is obtained
through
\begin{align}
\vec v^{(n+1/2)} &= \vec v^{(n)} + \vec a^{(n)} \Delta t_n / 2 , \\
\vec x^{(n+1)} &= \vec x^{(n)} + \vec v^{(n+1/2)} \Delta t_n , \\
\vec v^{(n+1)} &= \vec v^{(n+1/2)} + \vec a^{(n+1)} \Delta t_n / 2  .
\end{align}
We set the size $\Delta t_n$ of the timestep of step $n$ as 
\begin{align}
\Delta t_{n} = \min \left( \eta_{\rm orbit} \frac{V_{200}}{|\vec a^{(n)}|} ,
\eta_{\rm mesh} \frac{d_{\rm cell}^{(n)}}{|  \vec v^{(n)}|}
\right),
\end{align}
where $V_{200}$ is the circular velocity of the halo of the
constructed galaxy and $d_{\rm cell}^{(n)}$ is the dimension of the
mesh cell at the particle's current location. The dimensionless
coefficients $\eta_{\rm orbit}$ and $\eta_{\rm mesh}$ are meant to
ensure an accurate integration of the orbit and a precise accounting
of the time spent by the orbit in each of the bins used for recording
the density response.

We select the integrated timespan $T_i$ for each particle
individually. To this end we use the circular velocity at the
particle's initial position, and introduce a dimensionless factor
$\eta_{\rm timespan}$ for scaling the circular orbital time at the
local distance. Explicitly, we set
\begin{align}
T_{i} = \eta_{\rm timespan} \frac{2\pi |\vec{\hat x}_i|}{ v_{\rm
    circ}(\vec{\hat x}_i)},
\end{align}
where $v_{\rm circ}(\vec{\hat x}_i) \equiv (|\vec{\hat x}_i| \,
|\vec{\hat a}_i(\vec{\hat x}_i)|)^{1/2}$.  We typically found
$\eta_{\rm timespan}= 10.0$ to be sufficient, yielding an average
number of about 15 orbits for the particles of a typical halo.

\subsection{Optimization procedure} 

As discussed earlier, we in principle would like to optimize the
particles sequentially. Unfortunately, this immediately poses a
serious problem for any efficient parallelization.  If we enforce
strictly sequential iterative adjustments of the particle velocities
(such that a subsequent evaluation of the merit function already takes
the effects of a potential change of the previous particle's velocity
fully into account), then the optimization can evidently not be done
concurrently for several different particles.

However, we have found that in practice we still obtain good results
if we allow a small fraction of all particles to be treated
simultaneously, each remaining unaware of the changes in the other
particles until these are `committed' at the end of the concurrent
phase. With this approach, we can exploit massive parallelism in the
optimization procedure (as implemented in our {\small GALIC} code).

For definiteness, this practical aspect of our optimization scheme is
controlled by a parameter $f_{\rm opt}$ which gives the fraction of
particle orbits that can be set to new starting velocities without
taking note of each other. Our default values for this parameter is
$f_{\rm opt}=0.001$, meaning that our code will process the particles
in batches of size $f_{\rm opt} N_{\rm part}$ particles from a
randomly shuffled list of all particles. In each batch, all the trial
velocities are drawn and evaluated independently (hence this can be
done in parallel), and only at the end the velocity updates are
committed to the new global response of the system, affecting the next
batch.

When a particle is selected for optimization, we first randomly select
one of the three primary coordinate directions, and then replace the
corresponding velocity component with one drawn from the corresponding
Gaussian distribution. In this way, each of the optimizations
effectively couples only to one of the velocity dispersion measures.
We found this advantageous also for the following reason. To exclude
any possibility that systematic binning effects might prefer orbits
that start, for example, with positive $v_R$ as opposed to negative
$v_R$, we actually assess orbits by averaging the merit functions for
orbits both with $v_R$ and $-v_R$ velocities, and likewise for the
$v_z$ velocities.  This guarantees symmetry of the resulting velocity
distribution functions in these two directions, and in particular,
$\left<v_R\right> = \left<v_z\right> = 0$.  However, in the
$\phi$-direction, this reversal trick is not indicated, both because
here the symmetry of the binning procedure excludes the possibility of
any such effects by construction (unlike for the $R$- and
$z$-directions), and because in the $\phi$-direction orbits with a
reversed $\phi$-velocity are not necessarily equally probable.  We
note that to ensure that all particles remain bound, we reject any
trial velocity that is larger than $\eta_{\rm max} v_{\rm esc}$, where
$v_{\rm esc}$ is the local escape velocity and $\eta_{\rm max} =
0.9999$ is a parameter very close to 1.

In Figure~\ref{fig:convergence}, we show the decline of the value of
the merit function as a function of the number of velocity
optimizations that have been attempted by the code, in units of the
total particle number, for a typical initial conditions model where
the initial velocity guess were computed with the moment-based
method. We see that after $\sim 3$ optimization attempts for each
particle, the initial convergence speed slows down significantly, and
a stationary state in which no further improvement appears possible is
reached after approximately $\sim 5$ optimizations. We find this is a
quite typical behaviour in all of our models. Conservatively, we
usually run our models to $\sim 10$ optimizations per particle.

\subsection{Determination of the initial realization}

There are only two functions that need to be provided for any desired
density distribution that should be treated with our scheme.  For each
component of the system (i.e. halo, disk and/or bulge), one function
needs to return the density of the component at a given point, the
other must return a randomly sampled coordinate from the density field,
i.e.~the probability density of the corresponding point process must
be proportional to the density field. Having these functions in hand,
we can create the $\vec{\hat x}_i$ simply by randomly sampling each
density component present in the target system. Also, we can create a
(large) fiducial set of points for evaluating the force field to
arbitrary precision with a tree algorithm. Finally, we can make use of
the function returning the continuum density in solving the Jeans
equations.

Since in our approach we anyway compute the second moments with the
Jeans equations, we may as well initialize initial guesses for the
particle velocities $\vec{\hat v}_i$ by drawing randomly from Gaussian
distributions with the correct local dispersions. This corresponds to
the frequently invoked approximation of adopting triaxial Gaussians
for the local velocity distribution function, and since this is in
most cases reasonably close to the correct distribution, it
accelerates convergence. The iteration method is then in essence only
responsible for determining the higher-order moments of the velocity
distribution function.

\subsection{Parallelization approach}

Our C-code for creating stable initial conditions with the scheme
described here, {\small GALIC}, has been fully parallelized for
distributed memory machines using the message passing interface
(MPI). For calculating the gravitational field in the ($R,z$)-plane,
we let each MPI-task sample particles independently. The resulting
particle set is then subjected to a domain decomposition, and a
parallel distributed tree algorithm derived from the well-known
{\small GADGET} simulation code \citep{Springel2001, SpringelGadget2}
is invoked to compute the force field.

For creating the particles of the actual initial conditions, we again
let each MPI-task create a random, disjoint subset of the target
particle set for each mass component. Then we let each MPI-task work
independently and in parallel on the orbit optimizations associated
with one batch of size $f_{\rm opt} N_{\rm part}$.  The results are
then interchanged and the sums over the orbit responses are updated
accordingly, allowing the next cycle of optimizations to proceed. As a
result, the scalability of our code is essentially perfect provided
$f_{\rm opt}N_{\rm part}$ is substantially larger than $N_{\rm CPU}$,
otherwise work-load imbalances may become substantial as not all tasks
could then be expected to have roughly equal amounts of work in each
batch.

We have also made use of some of the I/O code from {\small GADGET}
when writing the final initial conditions to disk. They can be stored
in any of the three file formats supported by {\small GADGET}
(including one in HDF5-format), thereby simplifying the subsequent use
of the ICs with this simulation code, or the application of existing
file format conversion tools from {\small GADGET}'s format to other
simulation codes. Finally, our use of parallel I/O routines also
facilitates the creation of extremely large galaxy models, if desired.

\begin{figure}
\begin{center}
\setlength{\unitlength}{1cm}
\begin{picture}(10,6.8)
\resizebox{8.5cm}{!}{\includegraphics{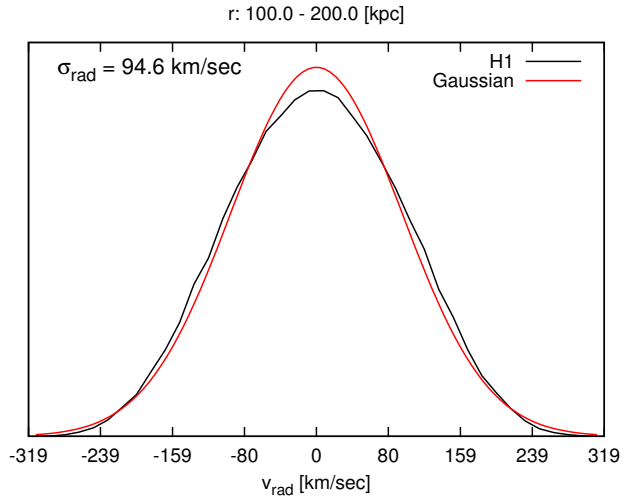}}
\end{picture}
\caption{Distribution function of the radial velocities in a Hernquist
  model within a thick radial shell. The black line gives the result
  of our code for the H1 model, while the red curve is a normal
  distribution with the same dispersion. The correct platykurtic shape
  of the distribution function (which is missed in moment-based
  approaches) is reproduced by our method.
\label{fig:CmpGaussian}}
\end{center}
\end{figure}

\begin{figure*}
\begin{center}
\resizebox{16cm}{!}{\includegraphics{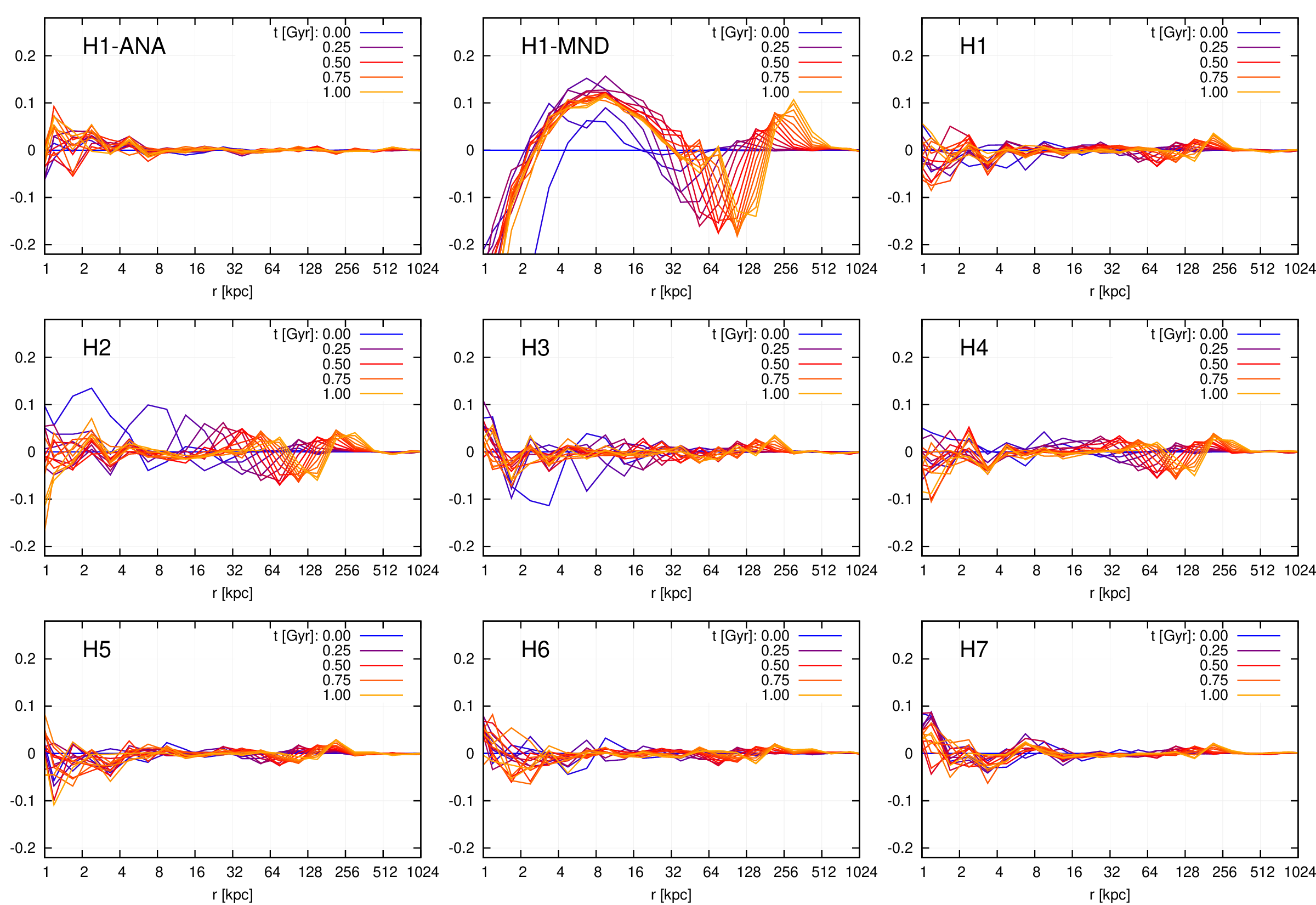}}
\caption{Density profile changes of different realizations of our
  halo-only H-models when evolved in time. The panels show the
  relative deviation of the spherically averaged density profile from
  the initial values at different radii and different simulation
  times. Different line colors mark the different times, as labelled.
  The top left panel corresponds to initial conditions for the
  isotropic H1-model realized with the analytic distribution function,
  while the top middle panel is for the moment-based approach, for
  comparison. All other results (H1 to H7) are for our new method as
  implemented in the {\small GALIC} code.
\label{fig:densHmodels}}
\end{center}
\end{figure*}

\section{Galaxy models}  \label{sec:galaxies}

The approach outlined above is quite general and can be used with
nearly arbitrary axisymmetric density profiles. For definiteness, we
describe in this section a specific set of parameterizations for dark
matter halos, stellar disks and stellar bulges, which we shall use in
our test galaxy models. These parameterizations follow models widely
employed in the literature.

We usually model the dark matter density profile as a spherically
symmetric halo with density
\begin{equation}
\rho_{\rm dm}(r) =
\frac{M_{\rm dm}}{2\pi}\,\frac{a}{r(r+a)^3},
\end{equation}
where $a$ is the scale factor. Following \citet{Springel2005}, we can
relate $a$ to the concentration $c$ of a corresponding NFW halo of
mass $M_{200} = M_{\rm dm}$ such that the shape of the density profile
in the inner regions is identical. The relation between $a$ and $c$ is
then given by
\begin{equation}
 a = \frac{r_{200}}{c}\sqrt{2 [ \ln(1+c) - c/(1+c)]},
\end{equation}
where $r_{200}$ and $M_{\rm 200}$ are the virial radius and virial
mass of the NFW halo, respectively.

We may also consider axisymmetric dark matter halos with either
prolate or oblate distortions. For simplicity, we assume that the
isodensity contours of the distorted shape are ellipses, effectively
created by linearly distorting the spherical shape along the symmetry
axis. If $s = a/c$ is the (radially constant) stretch factor, and
$a=b$ and $c$ are the axes of some isodensity ellipsoid, then a
prolate halo has $c/a > 1$ (and hence $s<1$), while an oblate halo has
$c/a < 1$ (with $s > 1$). We can then define the density profile of
the distorted halo as
\begin{equation}
\tilde{\rho}_{\rm dm}(R, z) \equiv s\,\rho_{\rm dm}(\sqrt{R^2 + s^2 z^2}),
\end{equation}
which leaves the total mass invariant.

For the disk, we adopt in general a model with exponential radial
scale length, and a ${\rm sech}^2$-profile in the vertical direction.
More specifically, the 3D disk density profile\footnote{We note that
  equation 10 of \cite{Springel2005} contains a typo in the form of an
  extraneous factor 1/2 in the argument of the sech-function. All
  model calculations in that paper have however been done correctly,
  based on equation 28 of \citet{Springel1999}, which is what we adopt
  here too.} is described by
\begin{equation}
\rho_\star(R,z) = \frac{M_{\star}}{4\pi z_0\, h^2}\, {\rm
sech}^2\left(\frac{z}{z_0}\right)\exp\left(-\frac{R}{h}\right).
\end{equation}
The disk scale length $h$ can either be set to a prescribed value, or
calculated by assuming that the disk contains a certain fraction of
the specific angular momentum of the halo \citep[see,
  e.g.,][]{Mo1998}.  We assume a radially constant scale height $z_0$
of the disk, but this could be easily modified if desired.  Usually,
we parameterize $z_0$ in terms of the disk scale length, with typical
disks lying in the range $z_0/h \sim 0.1-0.3$.

Finally, we model a stellar bulge (if present) with a Hernquist halo
as well, using the profile
\begin{equation} 
\rho_{\rm b}(r) = \frac{M_{\rm b}}{2\pi}\,\frac{b}{r(r+b)^3}
.  \end{equation} The bulge scale length $b$ is prescribed through a
parameter that gives its size in units of the halo's scale length.

We specify both the bulge and disk masses as fractions $m_{\rm d}$ and
$m_{\rm b}$ of the total mass, i.e.  $M_{\rm d}= m_{\rm d} M_{\rm
  tot}$ and $M_{\rm b}= m_{\rm b} M_{\rm tot}$. This parameterization
has previously been adopted in the study of \citet{Mo1998} on disc
structure, as well as in some earlier work on compound disc galaxy
models \citep[e.g.][]{Springel1999, Springel2005}.

\begin{figure}
\begin{center}
\resizebox{8.5cm}{!}{\includegraphics{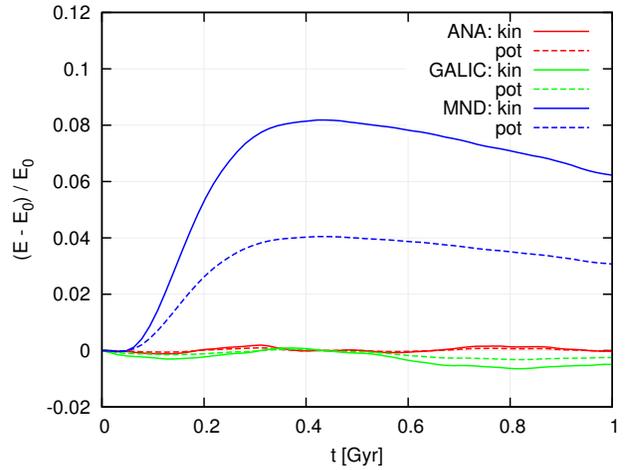}}
\caption{Time evolution of the relative change of kinetic and
  potential energies in the H1 model, for initial conditions realized
  either through the analytic distribution function (red), through the
  {\small GALIC} code (green), or with a moments based approach
  (blue).}
\label{fig:egyhqcmp}
\end{center}
\end{figure}

\begin{figure*}
\begin{center}
\resizebox{17.5cm}{!}{\includegraphics{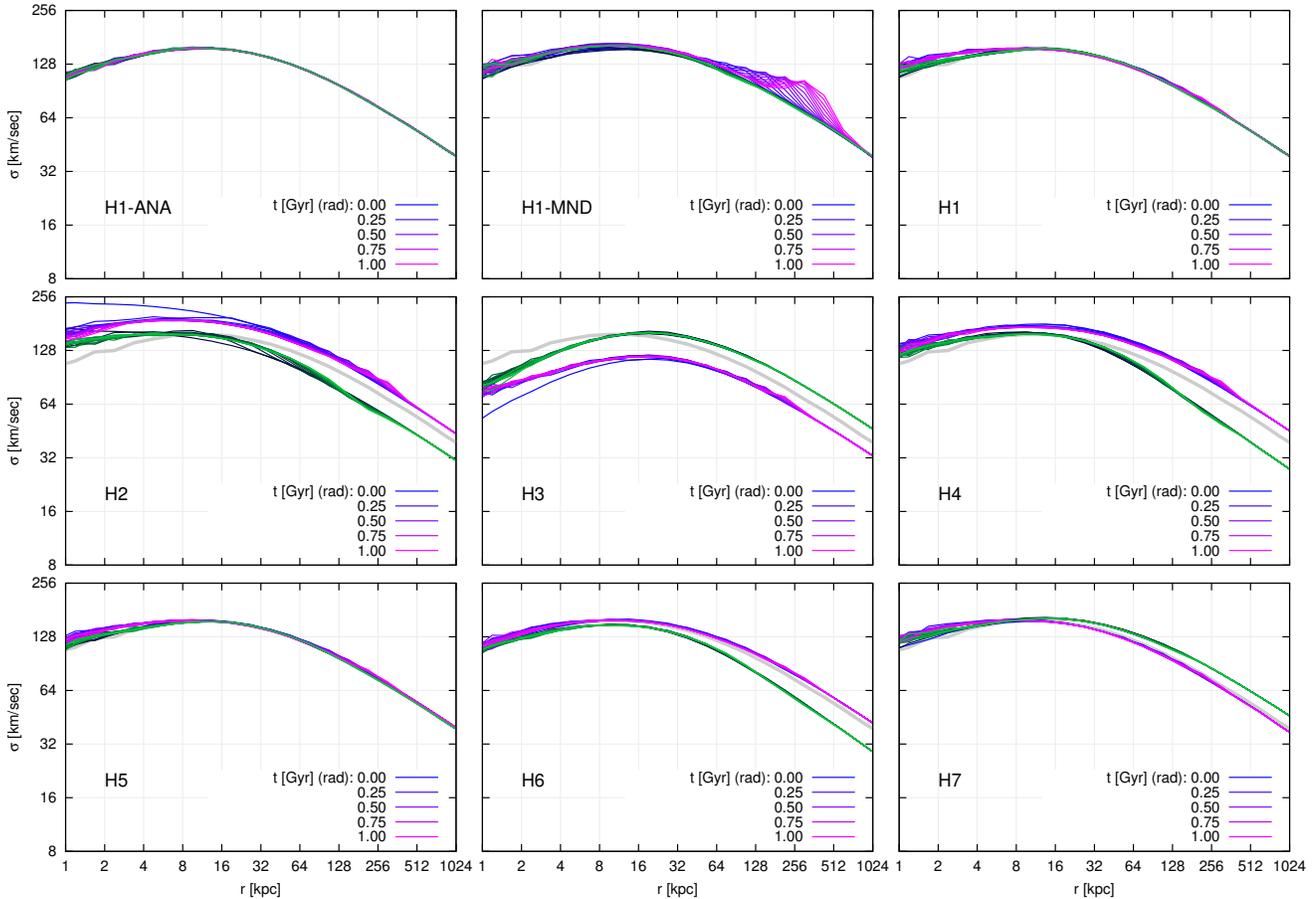}}
\caption{Radial and  azimuthal velocity dispersion profiles for our
  models H1 to H7. In each panel, we show the initial conditions
  produced by {\small GALIC}, and the evolved states after different
  times. The expected profiles based on the Jeans equations are shown
  underneath. The grey line reproduced the dispersion profile expected
  for the H1 model, for comparison. In the top row, we also show
  results for H1 obtained with the analytic distribution function (top
  left), and with the moment-based method (top, middle).}
\label{fig:velHmodels}
\end{center}
\end{figure*}

All the many reasonable combinations of the above components, together
with the various velocity structures possible for them, produce a
fairly large number of possibilities our code {\small GALIC} has to
deal with. In particular, requiring that a galaxy model always needs
to have a dark matter halo (either of spherical or oblate/prolate
shape), that a disk can either be present or absent, and that a bulge
is optional but may have different shapes if present (spherical,
oblate, or prolate), we already arrive at 12 possible combinations of
these three components. Of the corresponding models, only 2 have
spherically symmetric potentials (namely either the model with just a
spherical halo, or the model with spherical halo and a spherical
bulge), allowing ergodic $f(E)$ distribution functions or
$f(E,|\vec{L}|)$ models for them (besides axisymmetric $f(E,L_z)$ or
$f(E, L_z, I_3)$ distribution functions possible for all the
models). Allowing just for different combinations of these extra
velocity structures, this means that the 2 density models really
correspond to 6 possible variants. Similarly, the other 10 possible
density models give rise to 54 possible velocity variants, so that we
have of order 60 valid combinations of density model and associated
velocity structures. Of course, many of these models really correspond
to a continuum of further possibilities once the additional free
parameters describing, for example, the degree of net rotation or the
radial velocity anisotropy are used.

It is clear that we cannot present exhaustive tests of all these
possibilities in this work. Rather, we instead focus on a
representative selection of models which we list in
Table~\ref{TabSims}.  This sample of models covers a good fraction of
the space of possible model variants of interest, hence our tests
should give a good assessment of how well our techniques work in
practice. We consider, in particular, models that contain only a dark
matter halo (denoted as `H1', `H2', etc.), that feature a pure disk
embedded in a halo (labeled `D1', `D2', etc.), that contain a bulge
but no disk inside a halo (`B1', etc.), and finally, models that
feature both a disk and a bulge (`M1', `M2', etc.). In each of these
four groups, we consider models with a variety of velocity structures,
and/or different halo or bulge shapes. Detailed test results for the
produced initial conditions will be discussed in the next section.

\begin{figure}
\begin{center}
\resizebox{8.5cm}{!}{\includegraphics{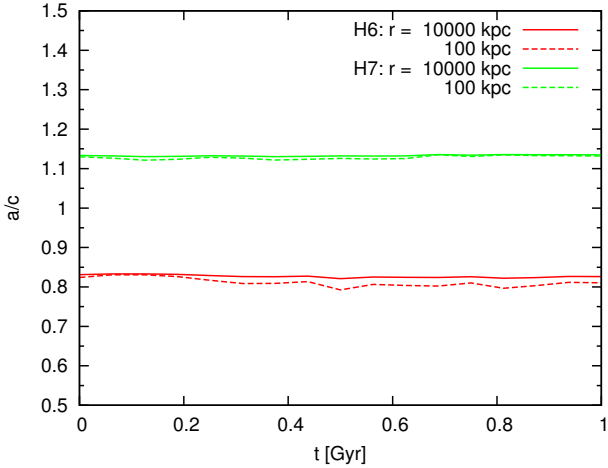}}
\caption{Halo shapes of models H6 and H7 as a function of evolution
  time.  We show the ratios of the principal eigenvalues of the 
moment-of-inertia tensor
  at two different radii as a function of time. The values reproduce 
the intended shapes according to Table~1, and are constant in time.
\label{fig:haloshapes}
}
\end{center}
\end{figure}

\begin{figure*}
\begin{center}
\resizebox{15cm}{!}{\includegraphics{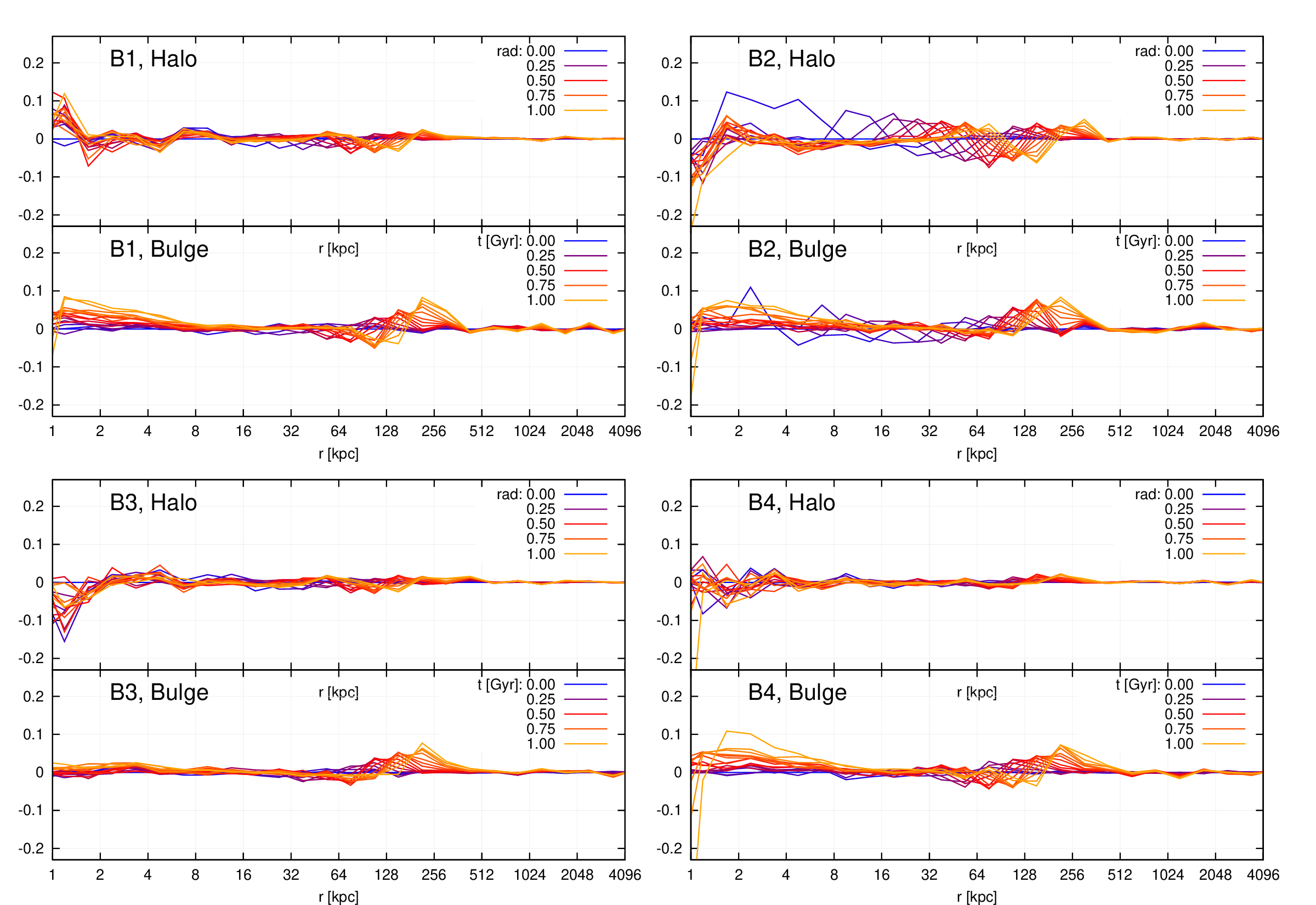}}
\caption{Density profiles changes of different bulge models (B-models
  in Table~1) when evolved in time.  For each of the four models, two
  panels are given that show the relative deviation of the spherically
  averaged density profile of halo and bulge components relative to
  the initial values, as a function of radius and for different
  simulation times. Different line colors mark the different times, as
  labelled.
\label{fig:densBmodels}}
\end{center}
\end{figure*}

\section{Test results}  \label{sec:testing}

\subsection{Models with a single halo component}

The simplest of our models is `H1', featuring a Hernquist dark matter
halo with an isotropic velocity distribution.  For definiteness, we
adopt $v_{200}=200\,{\rm km \, s^{-1}}$ and $c=10$ to set the total
mass and concentration of the halo, and we use $N=10^6$ particles in
order to have enough sampling points for a reliable measurement of the
produced velocity distribution function.

In Figure~\ref{fig:HqVelRadial}, we show radial and  azimuthal
velocity distribution functions measured from the ICs produced by our
code for this classic Hernquist model, where the analytic distribution
function is known analytically. We measure the distribution function
in a set of 6 radial shells, as labelled in the different panels. In
each panel, we show the analytic distribution function in blue, and
the one produced by the {\small GALIC} code in red (azimuthal
direction) and green (radial direction), respectively. We can nicely
see from the figure that the model calculated by {\small GALIC}
reproduces the expected distribution function rather well in all
radial shells, without any significant difference.  In particular,
note that the model produces the platykurtic nature of the velocity
distribution of the Hernquist sphere, which directly shows the
presence of higher-order moments that are missed by simpler
moment-based methods but are capture by our new approach. This is seen
explicitly in Figure~\ref{fig:CmpGaussian}, where we compare the
shape of the produced radial velocity distribution function to a
Gaussian with the same dispersion.

The most important critical test of a method's ability to create
initial conditions in equilibrium is however to check the stability of
the ICs in a self-consistent simulation under its own self-gravity.
To this end, we use the {\small GADGET} N-body code, with force
accuracy and time integration parameters set conservatively such that
energy conservation is excellent.  To control discreteness
  effects in the potential we set the gravitational softening length
  to a value of $0.05\,{\rm kpc}$. In this way we make sure that any
secular evolution that is seen really reflects imperfections of the
ICs rather than being influenced also by N-body integration errors
 or two-body relaxation.

Figure~\ref{fig:densHmodels} shows the relative deviation of the
spherically averaged density profile from the initial values at
different radii and different simulation times, for our H-models.
Different line colors mark the different times, as labelled.  We have
here restricted the simulation time to 1 Gyr, but note that nothing
qualitatively changes if this is expanded to 10 Gyrs, significantly
longer than the dynamical time of the galaxy model. Let us first focus
on a comparison of runs for three different initial conditions
constructed for H1, the isotropic Hernquist model.  The simulation
starting from ICs created with the analytic distribution function is
shown in the top left panel, the top right panel shows our {\small
  GALIC} technique, and the top middle panels gives the moment-based
method \citep[here realized with the {\small MAKENEWDISK} code
described in][]{Springel2005}.  As can be seen, our result (top right
panel) is nearly indistinguishable from the analytic initial
conditions.  There is a hint of some small deviations standing out of
the noise compared with the analytic solution, but this is very small
if real at all.  In contrast, for the moment-based method we see a
prominent perturbation propagating outwards, irreversibly changing the
mass distribution of the system as it relaxes to a new equilibrium
state.

Another view of this difference in the dynamical evolutions of these
three simulations is given in Figure~\ref{fig:egyhqcmp}, where we
compare the relative changes of the kinetic and potential energies of
the three runs as a function of time. We can see that both the N-body
realization drawn from the analytic Hernquist distribution function
and the {\small GALIC} result show essentially stationary energies
over the simulation, as expected from a virialized system in
equilibrium. In contrast, the moment-based approach shows a rapid
evolution in the energies in the first $\sim 300\,{\rm Myr}$, and only
then settles into a stationary state. Note that in this initial phase,
the central potential fluctuates, allowing individual particles to
change their energies and the system to relax to a new equilibrium.

We now turn to considerably more demanding models that have an
anisotropic, but still spherically symmetric velocity structure.
These are models H2, H3, and H4, characterized by asymmetry parameters
$\beta=-1$ (for H2) and $\beta = 0.5$ (for H3), corresponding to the
cases $\sigma_r^2 = \sigma_\theta^2 / 2$ and $\sigma_r^2 = 2
\sigma_\theta^2$, respectively. The model H4 adopts a radially varying
profile $\beta(r)$ as suggested by cosmological simulations.

\begin{figure}
\begin{center}
\resizebox{8.5cm}{!}{\includegraphics{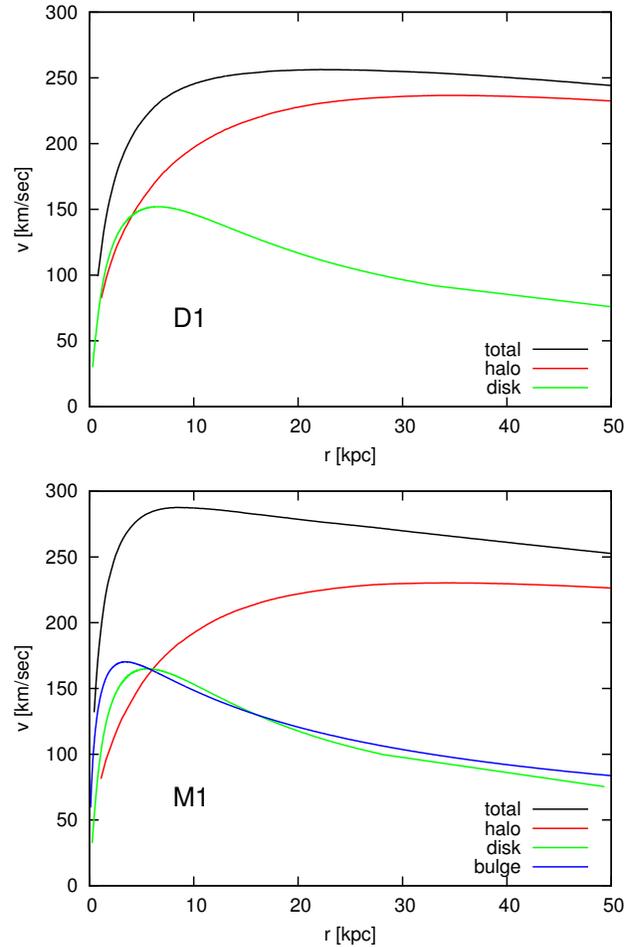}}
\caption{Rotation curves of the different mass components in our
  D-models (top panel) which contain only a dark matter halo and a
  stellar disk, and our M-models (bottom panel) which in addition
  contain a central bulge.
\label{fig:rotcurves}
}
\end{center}
\end{figure}

\begin{figure}
\begin{center}
\resizebox{8.5cm}{!}{\includegraphics{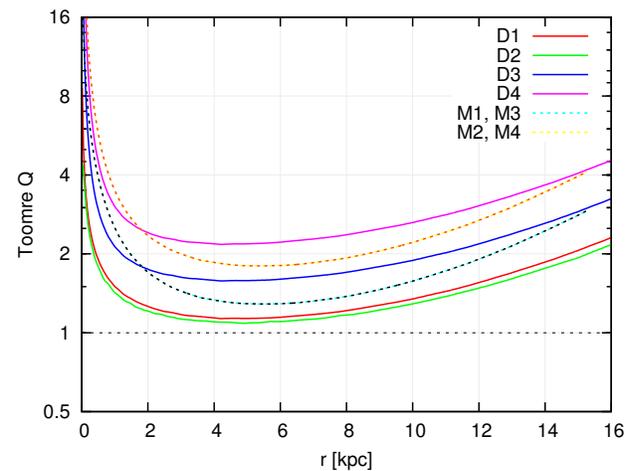}}
\caption{Stability against axisymmetric perturbations as expressed
  through the Toomre $Q$-parameter, for different D- and M-models.
\label{fig:toomreQ}
}
\end{center}
\end{figure}

In Figure~\ref{fig:velHmodels}, we show radial profiles of the radial
and  azimuthal velocity dispersion profiles for initial conditions
produced by {\small GALIC} for these 4 cases (and for completeness
also for all other H-models), both at the initial time and after
different times of evolution. For
reference, we also include in the figure panels the result for H1 (top
left panel) as a grey line, which is the isotropic $\beta=0$ case. We
see that the initial conditions code manages to accurately impose the
desired velocity anisotropy at the initial time. Upon time evolution,
these velocity dispersion profiles are quite well maintained, but not
perfectly in the very central regions for models H2, and to lesser
extent, H3. While these two models still manage to maintain a
directional difference in the velocity dispersions in the innermost
halo even after 1 Gyr of evolution, the initial profile is not fully
retained in the very central region. Overall, we consider these
results however still to be quite good.

\begin{figure*}
\begin{center}
\resizebox{18cm}{!}{\includegraphics{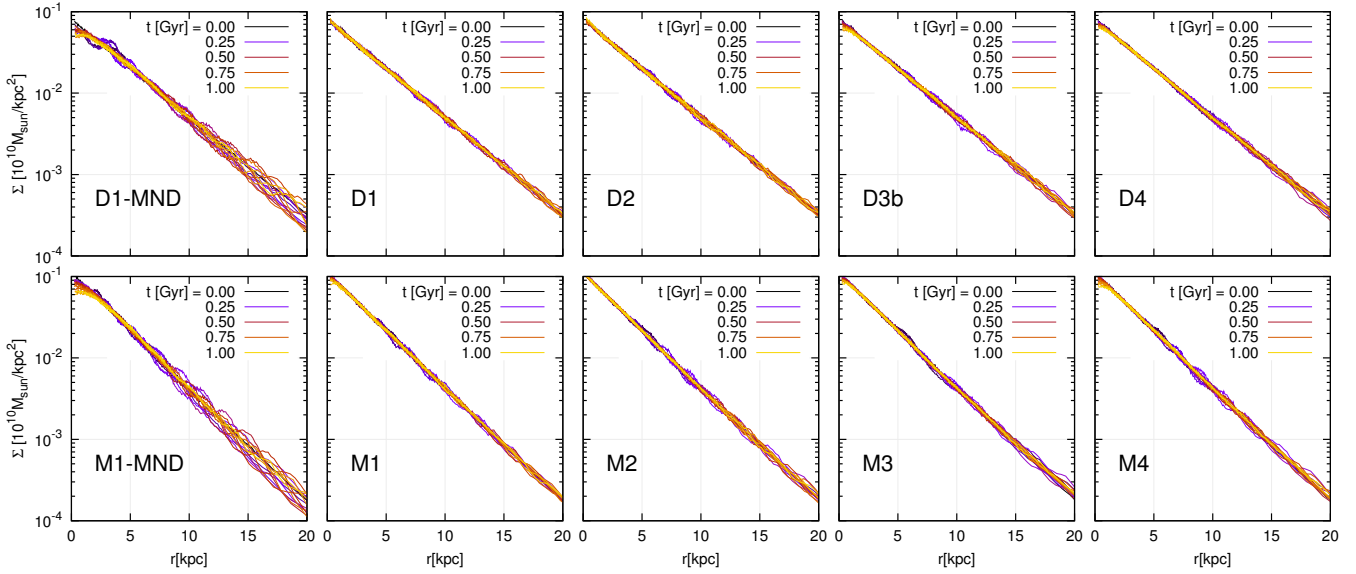}}
\caption{Radial surface density profiles of the stellar disk component
  under time evolution, for D1-D4 (top row), and M1-M4 (bottom
  row). The left column shows the D1 and M1 models again, but this
  time for initial conditions produced with the moment-based approach
  implemented in {\small MAKENEWDISK}.
  \label{fig:diskradprofiles}}
\end{center}
\end{figure*}

\begin{figure*}
\begin{center}
\resizebox{18cm}{!}{\includegraphics{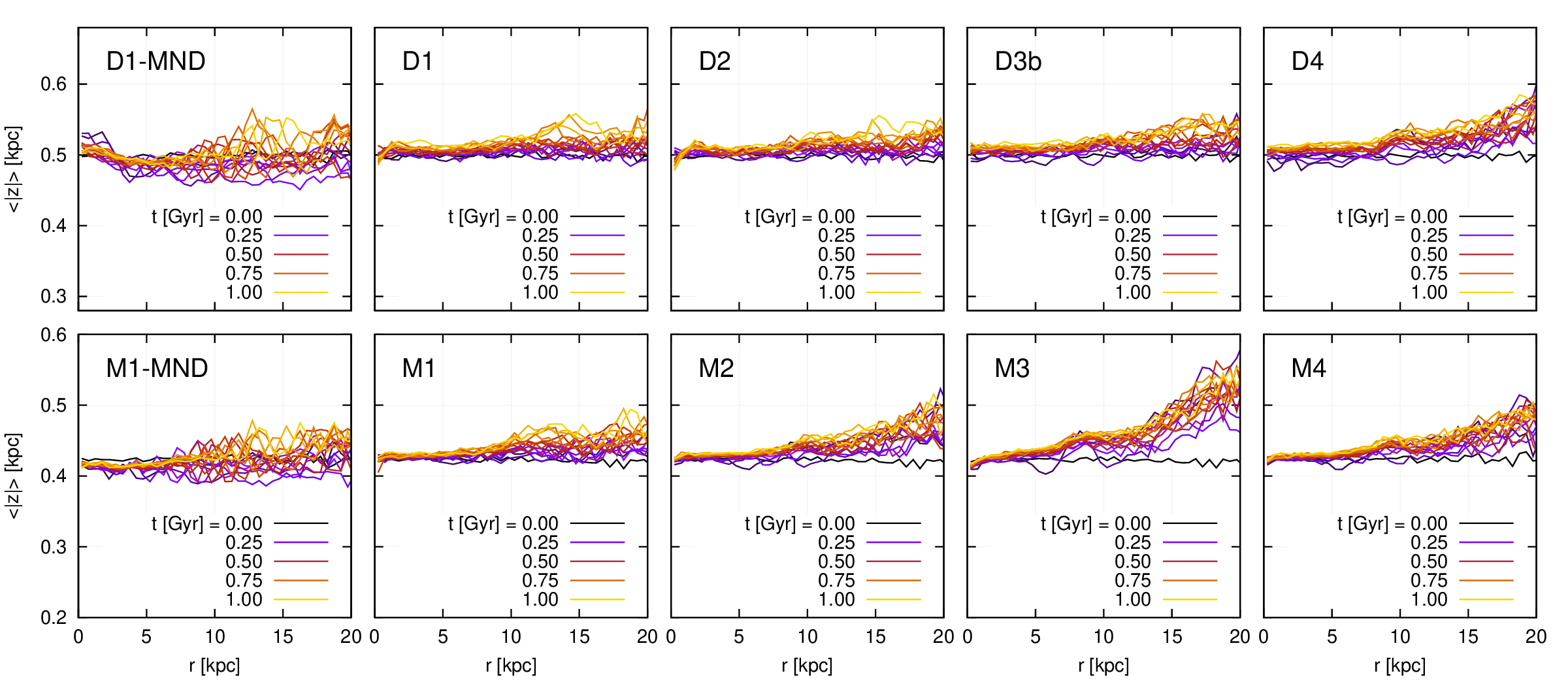}}
\caption{Vertical disk height as a function of radius at different
  times when our disk models are evolved in time, for the D1-D4 models
  (top row), and the M1-M4 models (bottom row).  In the left column,
  we show for comparison the corresponding results when the initial
  conditions for D1 and M1 are constructed with a moment-based
  approach.
   \label{fig:diskvertprofiles}}
\end{center}
\end{figure*}

We note however that the density profiles of these anisotropic models
are accurately retained even in the centres in these anisotropic
cases. Some of the panels of Figure~\ref{fig:densHmodels} report the
density variations of the anisotropic models H2-H4 upon time evolving
their ICs. The relative fluctuations of the density profiles are very
small, consistent with the findings for the simple H1 model. Note that
in Fig.~\ref{fig:densHmodels} we also include results for the models
H5, H6, and H7. These latter three models now feature an axisymmetric
assumption for their velocity structure. H5 is actually slowly
rotating, whereas H6 and H7 have prolate or oblate shape distortions,
respectively. The absence of any significant time evolution in the
spherically averaged density profiles shown in
Fig.~\ref{fig:densHmodels} indicates that these models are also rather
robust and in good equilibrium.

This is also confirmed by a look at their velocity dispersion profiles
shown in Figure~\ref{fig:velHmodels}, and a direct analysis of the
halos shapes of models H6 and H7 at the final times. Simple
measurements of the eigenvalues of their moment of inertia tensors as
a function of time (see Figure~\ref{fig:haloshapes}) confirm that the
imposed halo shapes are accurately retained over time. We also note
that the time evolution of the kinetic and potential energies (not
shown) confirms that the models are in good equilibrium.

\subsection{Systems with a bulge and a halo}

Next, we consider models that are slightly more complicated and
feature two different mass components of very different spatial
extent, a Hernquist halo with an embedded, much small stellar bulge,
also modelled with a Hernquist profile. Our model B1 simply consists
of two ergodic systems nested into each other. B2 varies that by
invoking different velocity anisotropies for halo and bulge, with a
preference for radial orbits in the halo and tangential ones in the
bulge. Finally, B3 and B4 test different shape distortions for halo
and bulge, under the assumption of an axisymmetric velocity structure
and no net rotation.

Figure~\ref{fig:densBmodels} shows the changes in the radially
averaged density profiles of these systems when they are evolved in
time, separately for the stellar bulge and the dark matter
components. The stability appears to be excellent in most cases,
something that is also confirmed by other measures, such as the time
evolution of kinetic and potential energies.  Only the B2
  model performs slightly worse, an outcome that we blame on the
  dominance of radial orbits in the dark matter component of this
  model even in the very centre (similar as in H2), and some of
  these orbits can be affected by the radial orbit instability
  \citep{Buyle2007}.

\subsection{Systems with a halo, a disk, and an (optional) bulge}

We now turn to the much more challenging case of models containing a
thin stellar disk. In Figure~\ref{fig:rotcurves}, we show the rotation
curves of our models containing just a halo and a disk (D-models, top
panel), and those of our models containing in addition a bulge as well
(M-models, bottom panel). In the innermost regions, the disk dominates
slightly over the spheroidal halo in the D-models. The specific
parameters chosen for the D-models are $m_d = 0.035$ and $\lambda =
0.035$, and the M-models contain additionally a bulge with $m_b=0.05$
and a scale length set to a tenth of that of the halo. We note that
these choices are somewhat arbitrary and not meant to represent a
specific system such as the Milky Way; our methods work with similar
quality when the parameters are varied over a plausible range.

Interestingly, depending on what assumptions we make about the
velocity structure of the disk systems, the expected stability with
respect to axisymmetric perturbations can be quite different. In
Figure~\ref{fig:toomreQ}, we show Toomre's $Q$-parameter for the disk
models D1 to D4, as well as for our M-models. As we see, D1 actually
nearly straddles the stability boundary at $Q=1$, and can hence be
expected to be somewhat more prone to axisymmetric perturbations than
D3, were $Q$ is boosted thanks to a higher radial velocity dispersion.

In Figure~\ref{fig:diskradprofiles}, we show the time evolution of the
azimuthally averaged projected disk surface density profiles, for
models D1-D4, and for M1-M4. We can see that all models are
reassuringly stable. The improvement compared with moment-based
methods such as that implemented the {\small MAKENEWDISK} code can
perhaps be best appreciated by comparing to the results for this
method, which are give in the leftmost panels of
Fig.~\ref{fig:diskradprofiles} for models D1 and M1.

Finally, a complementary view of the disk stability is obtained by
considering the time evolution of the vertical density structure of
the disks, wich is shown in Figure~\ref{fig:diskvertprofiles}. Again,
the models D1-D4 and M1-M4 are seen to retain their disk density
structure accurately, relatively independent of the different variants
of halo and bulge shapes, and the different degrees of rotation that
we tried. Only M3 performs noticeably worse than the other models in
the outer disk. When we compare the D1 and M1 models to corresponding
realizations obtained with the moment based approach (left most panels
in the figure), there is a clear improvement.

\subsection{Dependence on nuisance parameters}

Our iterative method for finding equilibrium galaxy models with the
{\small GALIC} code involves several free parameters, for example the
fraction of particles that is allowed to be concurrently optimized,
the number of optimization cycles before a randomization is carried
out, the resolution of the density response grid, the length of time
over which orbits are integrated, and a few more minor ones.

\begin{figure}
\begin{center}
\resizebox{8cm}{!}{\includegraphics{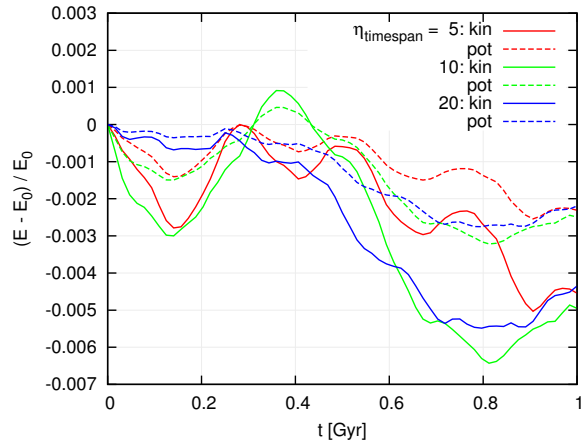}}
\caption{Test of the dependence of our results on the integration time
  of particle orbits. We show the relative changes in potential and
  kinetic energies when evolving initial conditions constructed with
  integration times lowered or increased by factors of two relative to
  our default value.  The particular system used here is M1, but
  similar results are found for other models and other changes of
  numerical parameters in our code.
\label{fig:changeinttime}
}
\end{center}
\end{figure}

We have carefully tested whether our results depend significantly on
the settings of any of these parameters. This is fortunately not the
case. We find that our results are rather robust when any of these
nuisance parameters is changed around our default settings. As a case
in point, we show in Figure~\ref{fig:changeinttime} an explicit test
for the number of orbits that are integrated, comparing results
produced for the M1 model where this parameter has been lowered by a
factor of 2, or increased by a factor of 2, compared with our default
choice. Reassuringly, we see that the density deviations occurring in
time evolutions of the produced ICs are of very similar magnitude,
i.e.~their quality appears indistinguishable.

We also find that that the grid resolution used for recording density
and velocity dispersion responses plays only a negligible role for the
results, provided the finest possible level is not overly coarse. This
can be understood as a result of our adaptive binning
prescription. Potentially more important may be the value of the
number of particles required in a cell before it may be split up in
finer cells. But we have also found here that varying this parameter
over a significant range does not change the results appreciably.
Finally, the last free parameter that we have extensively tested is
the integration accuracy of the orbits. Here our typical relative
energy errors for the integration of individual orbits are below
$10^{-3}$, already suggesting that this should be good enough and not
introduce any significant errors into the results. Indeed, investing
more computational effort and lowering the integration errors through
finder timestepping does not change our results in any significant
way.

\section{Discussion and Conclusions}  \label{sec:conclusion}

In this paper, we introduced a new iterative method for constructing
equilibrium N-body galaxy models.  The principle advantage of our
method is that it can produce N-body systems that are essentially in
exact equilibrium for rather general density distributions, making the
method ideal for studies of galaxy dynamics and numerical experiments
with isolated or colliding galaxies. Compared to alternative schemes
like the Schwarzschild method, our approach eliminates restrictions
arising from a finite orbit library or from required regularization
schemes.  Also, our method allows a natural inclusion of simulation
aspects like the need for a gravitational softening.

The test results we have analyzed show a considerable improvement of
the quality of the created initial conditions compared to existing
codes such as the moment-based {\small MAKENEWDISK}, a technique that
has been used in numerous studies over recent years.  This is possible
thanks to the absence of any assumptions in our approach with respect
to the importance of higher-order moments of the velocity distribution
function. The main disadvantage of our method lies in its higher
computational cost compared to moment-based approaches. However,
thanks to the scalable parallelization implemented of our code, this
should not be a serious restriction in practice. For
  example it took about 5 hours on 96 AMD-6174 cores (2.2 GHz) to
  compute high-quality solutions for our most compicated models M1-4,
  while for the one component models it took only 1 hour. And since
poor ICs may also impact any further scientific investigation, the
additional CPU effort invested for better ICs should in many cases be
well worth the effort.

Finally, we also note that numerous optimizations in our
code could well be made to reduce its CPU time consumption. For
example, the time to full convergence for a system with large $N$ may
be reduced considerably by first treating a smaller subsample of the
particles with correspondingly higher mass. Once this system has fully
converged, one could then create the large target realization from it
through bootstrap sampling, followed by briefly relaxing the big
system to the final equilibrium.

In future work, it might be interesting to extend our approach to
genuinely triaxial systems, which are of course considerably more
challenging than the axisymmetric case considered here. One could even
include additional phenomena such as figure rotation. An important
challenge is here to suitably store the density response grid. Here,
our approach, which only requires essentially one such response grid
should be considerably less restrictive than Schwarzschild's method
because the requirement to store a huge orbit library is avoided. In
the meantime, we publicly release our {\small GALIC} code 
  (http://www.h-its.org/tap/galic), hoping that it proves useful for
future N-body studies in galactic dynamics.

\section*{Acknowledgements}

We thank the referee for constructive comments that helped to improve
the paper. D.Y. and V.S.~acknowledge support by the DFG Research
Centre SFB-881 `The Milky Way System' through project A6.  This work
has also been supported by the European Research Council under ERC-StG
grant EXAGAL-308037 and by the Klaus Tschira Foundation.

\bibliographystyle{mn2e}
\bibliography{paper}

\begin{thebibliography}{}

\bibitem[\protect\citeauthoryear{{Athanassoula} \& {Misiriotis}}{{Athanassoula}
  \& {Misiriotis}}{2002}]{Athanassoula2002}
{Athanassoula} E.,  {Misiriotis} A.,  2002, \mnras, 330, 35

\bibitem[\protect\citeauthoryear{{Barnes} \& {Hernquist}}{{Barnes} \&
  {Hernquist}}{1992}]{Barnes1992}
{Barnes} J.~E.,  {Hernquist} L.,  1992, \araa, 30, 705

\bibitem[\protect\citeauthoryear{{Binney}, {Burnett}, {Kordopatis},
  {Steinmetz}, {Gilmore}, {Bienayme}, {Bland-Hawthorn}, {Famaey}, {Grebel},
  {Helmi}, {Navarro}, {Parker}, {Reid}, {Seabroke}, {Watson}, {Williams},
  {Wyse} \& {Zwitter}}{{Binney} et~al.}{2014}]{Binney2014}
{Binney} J.,  {Burnett} B.,  {Kordopatis} G.,  {Steinmetz} M.,  {Gilmore} G.,
  {Bienayme} O.,  {Bland-Hawthorn} J.,  {Famaey} B.,  {Grebel} E.~K.,  {Helmi}
  A.,  {Navarro} J.,  {Parker} Q.,  {Reid} W.~A.,  {Seabroke} G.,  {Watson} F.,
   {Williams} M.~E.~K.,  {Wyse} R.~F.~G.,    {Zwitter} T.,  2014, \mnras, 439,
  1231

\bibitem[\protect\citeauthoryear{{Binney} \& {Merrifield}}{{Binney} \&
  {Merrifield}}{1998}]{Binney1998}
{Binney} J.,  {Merrifield} M.,  1998, {Galactic Astronomy}

\bibitem[\protect\citeauthoryear{{Binney} \& {Tremaine}}{{Binney} \&
  {Tremaine}}{2008}]{BinneyTremaine2008}
{Binney} J.,  {Tremaine} S.,  2008, {Galactic Dynamics: Second Edition}.
Princeton University Press

\bibitem[\protect\citeauthoryear{{Buyle}, {van Hese}, {de Rijcke} \&
  {Dejonghe}}{{Buyle} et~al.}{2007}]{Buyle2007}
{Buyle} P.,  {van Hese} E.,  {de Rijcke} S.,    {Dejonghe} H.,  2007, \mnras,
  375, 1157

\bibitem[\protect\citeauthoryear{{Cretton}, {de Zeeuw}, {van der Marel} \&
  {Rix}}{{Cretton} et~al.}{1999}]{Cretton1999}
{Cretton} N.,  {de Zeeuw} P.~T.,  {van der Marel} R.~P.,    {Rix} H.-W.,  1999,
  \apjs, 124, 383

\bibitem[\protect\citeauthoryear{{Dehnen}}{{Dehnen}}{2009}]{Dehnen2009}
{Dehnen} W.,  2009, \mnras, 395, 1079

\bibitem[\protect\citeauthoryear{{D'Onghia}, {Springel}, {Hernquist} \&
  {Keres}}{{D'Onghia} et~al.}{2010}]{Donghia2010}
{D'Onghia} E.,  {Springel} V.,  {Hernquist} L.,    {Keres} D.,  2010, \apj,
  709, 1138

\bibitem[\protect\citeauthoryear{{D'Onghia}, {Vogelsberger} \&
  {Hernquist}}{{D'Onghia} et~al.}{2013}]{Donghia2013}
{D'Onghia} E.,  {Vogelsberger} M.,    {Hernquist} L.,  2013, \apj, 766, 34

\bibitem[\protect\citeauthoryear{{Hansen} \& {Moore}}{{Hansen} \&
  {Moore}}{2006}]{Hansen2006}
{Hansen} S.~H.,  {Moore} B.,  2006, \na, 11, 333

\bibitem[\protect\citeauthoryear{{Hernquist}}{{Hernquist}}{1990}]{Hernquist1990}
{Hernquist} L.,  1990, \apj, 356, 359

\bibitem[\protect\citeauthoryear{{Hernquist}}{{Hernquist}}{1993}]{Hernquist1993}
{Hernquist} L.,  1993, \apjs, 86, 389

\bibitem[\protect\citeauthoryear{{Hernquist} \& {Mihos}}{{Hernquist} \&
  {Mihos}}{1995}]{Hernquist1995}
{Hernquist} L.,  {Mihos} J.~C.,  1995, \apj, 448, 41

\bibitem[\protect\citeauthoryear{{Holley-Bockelmann}, {Mihos}, {Sigurdsson} \&
  {Hernquist}}{{Holley-Bockelmann} et~al.}{2001}]{Holley-Bockelmann2001}
{Holley-Bockelmann} K.,  {Mihos} J.~C.,  {Sigurdsson} S.,    {Hernquist} L.,
  2001, \apj, 549, 862

\bibitem[\protect\citeauthoryear{{Jaffe}}{{Jaffe}}{1983}]{Jaffe1983}
{Jaffe} W.,  1983, \mnras, 202, 995

\bibitem[\protect\citeauthoryear{{Jalali} \& {Tremaine}}{{Jalali} \&
  {Tremaine}}{2011}]{Jalali2011}
{Jalali} M.~A.,  {Tremaine} S.,  2011, \mnras, 410, 2003

\bibitem[\protect\citeauthoryear{{Kazantzidis}, {Bullock}, {Zentner},
  {Kravtsov} \& {Moustakas}}{{Kazantzidis} et~al.}{2008}]{KazantzidisI2008}
{Kazantzidis} S.,  {Bullock} J.~S.,  {Zentner} A.~R.,  {Kravtsov} A.~V.,
  {Moustakas} L.~A.,  2008, \apj, 688, 254

\bibitem[\protect\citeauthoryear{{Kuijken} \& {Dubinski}}{{Kuijken} \&
  {Dubinski}}{1995}]{Kuijken1995}
{Kuijken} K.,  {Dubinski} J.,  1995, \mnras, 277, 1341

\bibitem[\protect\citeauthoryear{{Merritt}}{{Merritt}}{1985}]{Merritt1985}
{Merritt} D.,  1985, \mnras, 214, 25P

\bibitem[\protect\citeauthoryear{{Mo}, {Mao} \& {White}}{{Mo}
  et~al.}{1998}]{Mo1998}
{Mo} H.~J.,  {Mao} S.,    {White} S.~D.~M.,  1998, \mnras, 295, 319

\bibitem[\protect\citeauthoryear{{Navarro}, {Frenk} \& {White}}{{Navarro}
  et~al.}{1997}]{Navarro1997}
{Navarro} J.~F.,  {Frenk} C.~S.,    {White} S.~D.~M.,  1997, \apj, 490, 493

\bibitem[\protect\citeauthoryear{{Osipkov}}{{Osipkov}}{1979}]{Osipkov1979}
{Osipkov} L.~P.,  1979, Soviet Astronomy Letters, 5, 42

\bibitem[\protect\citeauthoryear{{Rodionov}, {Athanassoula} \&
  {Sotnikova}}{{Rodionov} et~al.}{2009}]{Rodionov2009}
{Rodionov} S.~A.,  {Athanassoula} E.,    {Sotnikova} N.~Y.,  2009, \mnras, 392,
  904

\bibitem[\protect\citeauthoryear{{Satoh}}{{Satoh}}{1980}]{Satoh1980}
{Satoh} C.,  1980, \pasj, 32, 41

\bibitem[\protect\citeauthoryear{{Schwarzschild}}{{Schwarzschild}}{1979}]{Schwarzschild1979}
{Schwarzschild} M.,  1979, \apj, 232, 236

\bibitem[\protect\citeauthoryear{{Sellwood} \& {Binney}}{{Sellwood} \&
  {Binney}}{2002}]{Sellwood2002}
{Sellwood} J.~A.,  {Binney} J.~J.,  2002, \mnras, 336, 785

\bibitem[\protect\citeauthoryear{{Siebert}, {Bienaym{\'e}}, {Binney},
  {Bland-Hawthorn}, {Campbell}, {Freeman}, {Gibson}, {Gilmore}, {Grebel},
  {Helmi}, {Munari} \& {Navarro}}{{Siebert} et~al.}{2008}]{Siebert2008}
{Siebert} A.,  {Bienaym{\'e}} O.,  {Binney} J.,  {Bland-Hawthorn} J.,
  {Campbell} R.,  {Freeman} K.~C.,  {Gibson} B.~K.,  {Gilmore} G.,  {Grebel}
  E.~K.,  {Helmi} A.,  {Munari} U.,    {Navarro} J.~F.,  2008, \mnras, 391, 793

\bibitem[\protect\citeauthoryear{{Springel}}{{Springel}}{2005}]{SpringelGadget2}
{Springel} V.,  2005, \mnras, 364, 1105

\bibitem[\protect\citeauthoryear{{Springel}, {Di Matteo} \&
  {Hernquist}}{{Springel} et~al.}{2005}]{Springel2005}
{Springel} V.,  {Di Matteo} T.,    {Hernquist} L.,  2005, \mnras, 361, 776

\bibitem[\protect\citeauthoryear{{Springel} \& {White}}{{Springel} \&
  {White}}{1999}]{Springel1999}
{Springel} V.,  {White} S.~D.~M.,  1999, \mnras, 307, 162

\bibitem[\protect\citeauthoryear{{Springel}, {Yoshida} \& {White}}{{Springel}
  et~al.}{2001}]{Springel2001}
{Springel} V.,  {Yoshida} N.,    {White} S.~D.~M.,  2001, \na, 6, 79

\bibitem[\protect\citeauthoryear{{Syer} \& {Tremaine}}{{Syer} \&
  {Tremaine}}{1996}]{Syer1996}
{Syer} D.,  {Tremaine} S.,  1996, \mnras, 282, 223

\bibitem[\protect\citeauthoryear{{van den Bosch}, {Gebhardt}, {G{\"u}ltekin},
  {van de Ven}, {van der Wel} \& {Walsh}}{{van den Bosch}
  et~al.}{2012}]{Bosch2012}
{van den Bosch} R.~C.~E.,  {Gebhardt} K.,  {G{\"u}ltekin} K.,  {van de Ven} G.,
   {van der Wel} A.,    {Walsh} J.~L.,  2012, \nat, 491, 729

\bibitem[\protect\citeauthoryear{{Vandervoort}}{{Vandervoort}}{1984}]{Vandervoort1984}
{Vandervoort} P.~O.,  1984, \apj, 287, 475

\bibitem[\protect\citeauthoryear{{Widrow} \& {Dubinski}}{{Widrow} \&
  {Dubinski}}{2005}]{Widrow2005}
{Widrow} L.~M.,  {Dubinski} J.,  2005, \apj, 631, 838

\end{thebibliography}

\label{lastpage}

\end{document}